\documentclass{aa}  

\usepackage{graphicx}
\usepackage{xcolor}
\usepackage{txfonts}
\usepackage{url}
\usepackage{natbib}
\usepackage{verbatim}
\usepackage{amsmath}

\def\beq{\begin{equation}}
\def\eeq{\end{equation}}
\def\xbh{x_{\rm BH}}
\def\ybh{y_{\rm BH}}
\def\zbh{z_{\rm BH}}
\def\xorb{x_{\rm orb}}
\def\yorb{y_{\rm orb}}
\def\zorb{z_{\rm orb}}
\def\ibh{i_{\rm BH}}
\def\obh{\omega_{\rm BH}}
\def\iorb{i_{\rm obs}}
\def\oorb{\omega_{\rm obs}}
\def\Oorb{\Omega_{\rm obs}}
\def\Obh{\Omega_{\rm BH}}
\def\psibh{\psi_{\rm BH}}
\def\xobs{x_{\rm obs}}
\def\yobs{y_{\rm obs}}
\def\zobs{z_{\rm obs}}
\def\sgra{SgrA$^*\,$}
\def\sdisk{$\mathcal{S}_{\rm disk}\,$}
\def\rdisk{$\mathcal{R}_{\rm disk}\,$}
\def\wdisk{$\mathcal{W}_{\rm disk}\,$}
\usepackage{keyval}
\usepackage{amsmath}

\begin{document}

   \title{Effects of a disk structure on stellar motion at the Galactic Center}

   \author{A. Foschi
          \inst{1}, F. H. Vincent \inst{1}, T. Paumard \inst{1} \and G. Perrin \inst{1} 
          }

   \institute{LIRA, Observatoire de Paris - PSL, 5 Pl. Jules Janssen, 92190 Meudon, France} 
 
\abstract
  {Stellar orbits are key for probing the environment of the supermassive black hole at the Galactic Center, Sagittarius A$^*$. The impact on these orbits of an extended mass distribution must be disentangled from that of the central compact object. So far, such mass has been assumed to be spherically distributed, for simplicity. However, there are reasons to believe that the extended mass component should rather be flattened, creating disk-like structures.} 
  {We investigate the effects that a thin disk structure would have on stars at the Galactic Center, focusing on star S$2$ and S$301$. To simulate a realistic scenario we assume that the disk coincides with the clockwise stellar disk observed at the Galactic Center. } 
  {We derive analytically the acceleration exerted by a disk with power law surface density $\Sigma \propto r^{-\gamma}$. We use this acceleration to compute the osculating equations and the variations of the orbital elements over one orbit, showing how the latter depend on the orientation of the disk with respect to the orbital plane. We derive these variations for both S$2$ and S$301$ in the specific case of the clockwise disk.} 
  {We find that the disk structure induces a secular shift in the semi-latus rectum, an extra in-plane precession and an out-of-plane precession. The former is neither present at the low-order post Newtonian description that we use for the black hole, nor when a spherical mass distribution is considered. The latter can be competitive with the same effect due to the Lense-Thirring precession induced by the spin of Sagittarius A$^*$ on S$301$ motion, depending on the mass, the radial extent and the orientation of the disk. For S$2$, considering a reasonable value of the disk mass, we show that the secular effect on the variation of the semi-latus rectum, $\Delta p_{\rm star} \sim10 \, \mu \rm as$, is commensurate with the typical astrometric precision currently achievable on stellar orbits. The typical in-plane pericenter shift due to the disk, $\Delta \oorb \sim 0.2^{\circ}$, is within a factor of a few similar to that of a Plummer distribution, however, specific configurations with a ten times more massive disk than the Plummer distribution are shown to lead to similar secular trends, showing that a disk structure might accommodate more mass. The typical out-of-plane precession $\Delta \Theta \sim 0.03^{\circ}$ is comparable with the  current measurements precision and thus useful to constrain the disk mass. For S$301$, the out-of-plane disk effect is a factor of ten smaller than the spin effect, assuming the most favourable spin configuration. If that assumption is dropped, however, the disk precession can be substantially larger than the spin precession, showing the importance of constraining an extended disk component.}
  {Since the Lense-Thirring precession is negligible in S$2$ motion, the out-of-plane precession induced by the disk can be used to place upper limits on the non luminous mass in the clockwise stellar disk or other disk-like structures at the Galactic Center. The limits might significantly differ (by a factor of up to an order of magnitude) from those obtained for spherical distributions and dependent on the disk parameters. Our results show the key importance of constraining an extended disk structure for monitoring stellar orbits in view of constraining the properties of the central compact object, in particular its spin. Once the mass estimates are at hand, one can quantify the effect that such a disk may have on S$301$ motion and thus the degeneracy with the future measurement of Sagittarius A$^*$ spin. We check that the presence of other structures (the giant molecular cloud and the circumnuclear disk) located outside the S-cluster range has negligible effects on both stars.}


\maketitle
\authorrunning
\titlerunning

\section{Introduction}

Most of the works on stellar dynamics at the Galactic Center (GC) are performed assuming a spherically symmetric distribution of matter around the supermassive black hole (SMBH) Sagittarius A$^*$ (\sgra), mainly to reproduce dark matter models \citep{Zakharov:2007fj, Lacroix:2018zmg,Heissel:2021pcw, Shen:2023kkm, Lechien:2023psa, Paul:2026}. 

For the most known profiles, i.e., Plummer and cusp, the current upper limit derived by the GRAVITY Collaboration using multiple S-stars suggests an extended mass $M_{\rm ext} < 1200 M_{\odot}$ at $1\sigma$ confidence level, within $r_0 \sim 0.01\, \rm pc$ from \sgra \citep{GRAVITY:2021xju,GRAVITY:2024tth}.

However, observations at the GC have shown the presence of a population of young stars in a coherent disk-like structure orbiting \sgra, known as clockwise (CW) disk, since it moves clockwise with respect to the Earth \citep{Genzel:2003cn, Levin:2003, Alexander:2005jz, Paumard:2006, Bartko:2008ad, Lu:2008iz, vonFellenberg:2022lyo}. 


The existence of such disk-like structures around \sgra can be used to put constraints on the vector resonant relaxation process in stellar systems \citep{Kocsis:2011, Fouvry:2019} and on the presence of intermediate mass black holes (BHs) at the GC \citep{Fouvry:2023}.

The same process has been used to show that stellar and intermediate mass BHs in the vicinity of \sgra can eventually form a warped disk, which can align with the CW disk and even provide an explanation for its origin if the right initial conditions are fulfilled  \citep{Szolgyen:2018zra, Tremaine:2020, Szolgyen:2021, Mathe:2022azz}. 

Moreover, it is well known that matter that accretes the central BH does it in a disk like structure, even if the current estimate for \sgra accretion flow seems to suggest a very low accreting mass rate \citep{Narayan:1995dn, Ressler:2018yhi}.

For this reason, analysing the effects that a planar, axisymmetric distribution of matter would have on the orbital motion of S-stars can provide meaningful insights on \sgra surroundings. 

In this paper we consider the effect of a thin disk on the orbital motion of a star like S$2$ or S$301$, using a pure Newtonian framework. The latter is justified since S-stars orbit sufficiently far from \sgra to be in almost Keplerian orbits. 

However, to take into account the relativistic nature of \sgra, we also include the Schwarzschild precession via the Post Newtonian (PN) formalism \citep{PoissonWill2012,Will:2014kxa}, since the latter has been detected in S$2$ motion at $10\sigma$ confidence level \citep{GRAVITY:2024tth}.  

Moreover, as we will see, the presence of a disk induces an out-of-plane precession that may be comparable in magnitude with the Lense-Thirring (LT) precession \citep{Misner:1973prb}, i.e. the frame dragging effect exerted by \sgra rotation, which is modelled via the $1.5$ PN equations of motion \citep{Dayem:2025cki}. 

Since the newly discovered star S$301$ \citep{GRAVITY:2026} will make possible the measurement of \sgra spin within a few years, we also determine how these two effects compare and the possible degeneracy between the two, with a focus on the CW disk configuration as a realistic example. 

Since with the latter analysis we moved further away than the usual S-cluster scale, we briefly show that the effect of other structures located within the innermost parsec of \sgra have in fact negligible effects.

\section{Methods}

As a first step, we consider a non-rotating SMBH with mass $M_{\bullet}$. The effect of the black hole on stellar orbits at the GC will be modeled by means of a first order PN expression in terms of harmonic coordinates $(t, r, \theta, \phi)$. 
The BH reference frame is identified by $\{x_{\rm BH},y_{\rm BH}, z_{\rm BH}\}$, with the disk lying in its equatorial plane ($z_{\rm BH} = 0$). 

\subsection{Setup and reference frames}
\label{sec:setup}
The system we consider is made up of three different reference frames, all of which are direct orthonormal triads centered at the BH position.

The first is the BH frame which, in what follows, corresponds to the disk frame.

The second frame is the orbital frame of the star defined by $\{x_{\rm orb}, y_{\rm orb}, z_{\rm orb}\}$, such that $x_{\rm orb}$ points towards the direction of the pericenter and $z_{\rm orb}$ is aligned with the direction of the angular momentum of the orbit. 

We can also define a time-dependent frame that comoves with the star and it is identified by the triad $\{\mathbf{n}, \boldsymbol{\lambda}, \mathbf{z}\}$ also known as Gaussian frame (e.g. \cite{PoissonWill2012}), where $\mathbf{n}$
points from the BH to the star position in the orbital plane and $\mathbf{z} = \mathbf{\zorb}$. 

The BH equatorial plane and the orbital plane intersect in correspondence of the line of nodes, which is directed by the vector $\boldsymbol{\ell}_{\rm BH}$, pointing from the origin of the BH frame towards the ascending node of the orbit (where we remind that the ascending node of the orbit corresponds to the node where the $z_{\rm BH}$ coordinate of the star switches from negative to positive). 

The BH and orbit frames are related by the three angles $\ibh$ (inclination from $\zbh$ to $\zorb$),  $\omega_{\rm BH}$ (angle from $\boldsymbol{\ell}_{\rm BH}$ to $\xorb$, positively counted in the direction of stellar motion and lying in the orbital plane) and $\Omega_{\rm BH}$ (angle from $\xbh$ to $\boldsymbol{\ell}_{\rm BH}$, lying in the equatorial plane). The direction of positive $\Omega_{\rm BH}$ is not relevant, as the choice of this parameter has no consequence in the dynamics. 


The third reference frame is the observer one $\{\xobs, \yobs, \zobs\}$, still centered at the BH location, that represents the sky frame where data are collected. In our setup $\xobs = \text{DEC}$ (declination, North), $\yobs= \text{R.A.}$ (right ascension, East) and $\zobs$ points from the BH towards the Earth. Note that this convention is different from that used in GRAVITY observational papers, where $\zobs$ is defined from Earth towards the BH, and it is the same as \cite{Heissel:2021pcw}.

The relative inclination and orientation of the orbital frame with respect to the observer frame is encoded in the three angles $\iorb$, $\oorb$ and $\Oorb$.

The corresponding line of nodes (intersection between orbit and sky planes) is directed by $\boldsymbol{\ell}_{\rm obs}$ pointing from the BH towards the ascending node, defined similarly as above by the point of the orbit where the sign of the star's $\zorb$ coordinate switches from negative to positive. The vector $\boldsymbol{\ell}_{\rm obs}$ clearly differs from $\boldsymbol{\ell}_{\rm BH}$ defined above. 

$\iorb$ is inclination between $\zobs$ and $\zorb$, $\oorb$ is the angle between $\boldsymbol{\ell}_{\rm obs}$ and $\xorb$, lying in the orbit plane and positively counted in the direction of stellar motion, while $\Oorb$ is the angle between $\xobs$ and $\boldsymbol{\ell}_{\rm obs}$, lying in the plane of the sky and positively counted East of North.

\begin{figure}
   \centering
   \includegraphics[width=\columnwidth]{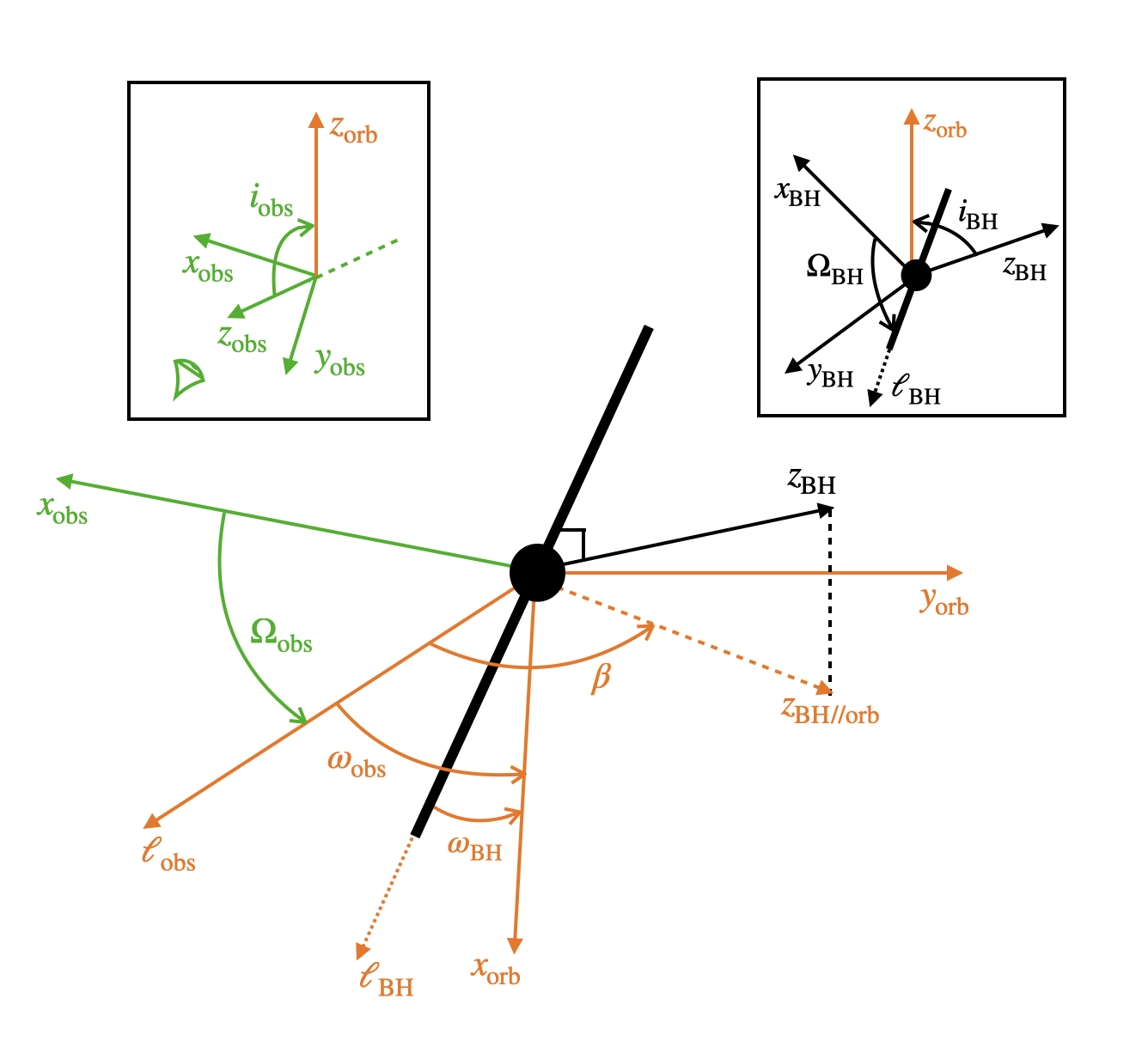}
      \caption{Schematic representation of the three reference frames involved. The orbital plane $(\xorb, \yorb)$ is in orange and angles defined in this plane are the same colour. In green there is the observer frame and the angles defined there (left inset). Finally, in black there is the BH/disk frame, with relative angles (right inset). }
         \label{fig:full_setup}
\end{figure}

Moreover, as it is shown in Appendix~\ref{app:omega},  $\obh$ can be related to $\oorb$ via the following relation
\beq
\omega_{\rm BH} = \frac{\pi}{2} + \omega_{\rm orb} -\beta\, ,
\label{omega_bh}
\eeq

where $\beta$ represents the angle between $\ell_{\rm obs}$ and $z_{\rm BH//orb}$, i.e., $\zbh$ projected onto the orbital frame (for a schematic representation of the setup with relative angles, see Figure~\ref{fig:full_setup}).

\subsection{Acceleration due to a planar distribution of matter}
In this setup, the star feels a total acceleration given by:
\begin{equation}
    \mathbf{a} = - \frac{G M_{\bullet}}{r^3} \mathbf{r} + \mathbf{a}_{\rm disk} + \mathbf{a}_{\rm 1PN}\, ,
\end{equation}
where $M_{\bullet}$ is the SMBH mass and $\mathbf{a}_{\rm 1PN}$ is the first PN correction to the equations of motion and can be expressed in harmonic coordinates as
\begin{equation}
    \mathbf{a}_{1 \rm PN} = \frac{ G M_{\bullet}}{c^2 r^2} \left[\left(\frac{4 G M_{\bullet}}{r} - v^2\right) \frac{\boldsymbol{r}}{r} + 4 \dot{r}\boldsymbol{v}  \right] \, ,
\label{1pn}
\end{equation}
where the dot represents a derivative with respect to coordinate time, while $\mathbf{a}_{\rm disk}$ is the acceleration due to the disk. The latter is obtained in a pure Newtonian framework.

\subsubsection{Single uniform one-dimensional ring}
\label{subsec:uniform_ring}

We start with the simplest case of a zero-thickness ring with radius $R$ and uniform linear density $\lambda = M_{\rm ring}/2 \pi R$ and a point mass $D$ in its gravitational potential.  The Cartesian coordinates of $D$ are $(x,y,z)$, while its cylindrical coordinates are $(d, \phi_D, z)$.

\begin{figure}
   \centering  \includegraphics[width=\hsize]{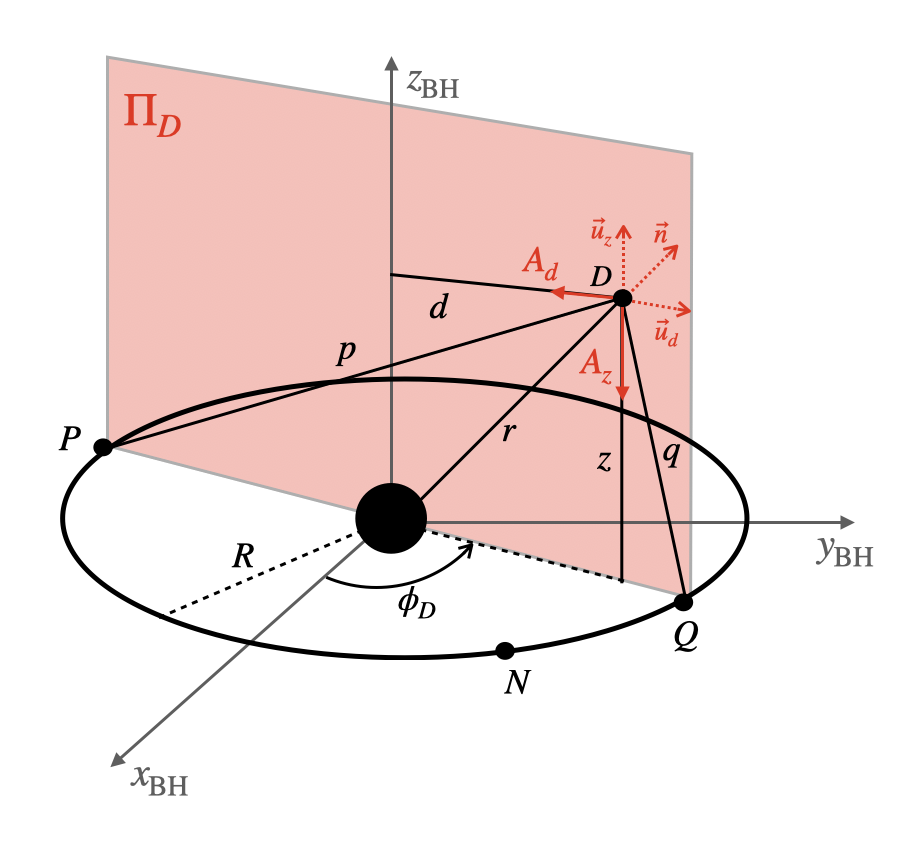}
      \caption{Schematic representation of the disk setup with parallel and perpendicular components of the acceleration in Eq.~\eqref{Ap}-\eqref{Aq}. We note that the three unit vectors all lie on the red plane $\Pi_D$, with $\vec{n}$ being the unit vector along the three-dimensional radius $r$.}
         \label{fig:disk}
\end{figure}

It is clear by symmetry that the force due to the ring felt by $D$ must lie in the plane $\Pi_D$ containing $D$, the central BH and the normal vector to the ring (see Figure \ref{fig:disk}). 

This plane crosses the ring in two points, namely $P$ and $Q$, which correspond to the closest and furthest neighbours of $D$ along the ring. It is thus natural to compute the location of these two points.

Let us take a generic point $N$ that lies on the ring, with cylindrical coordinates $(R,\phi_N, 0)$. The cylindrical distance from point $D$ to the center where the BH is located is $d = \sqrt{\xbh^2 + \ybh^2}$, so the distance between $D$ and $N$ can be written as
\beq
\begin{split}
|DN|^2 =& (d \cos \phi_D - R \cos \phi_N)^2 + (d \sin \phi_D - R \sin \phi_N)^2 + z^2\\
= & d^2 +R^2 + z^2 - 2 d R (\cos \phi_D \cos \phi_N + \sin \phi_D \sin \phi_N )\, .
\end{split}
\eeq
The extrema of $|DN|^2$ are easy to find in terms of the azimuthal angle $\phi$, and are given by 
\beq
\tan \phi = \tan \phi_D \, \,\, \Rightarrow \phi = \phi_D \pm n \pi\, . 
\eeq 

When $\phi = \phi_D$ ($\phi = \phi_D+\pi$) the distance is minimum (maximum) and corresponds to the distance $|DQ|$ ($|DP|$). We can write these distances as
\begin{equation}
\begin{split}
    p =\sqrt{(d + R)^2 + \zbh^2},\,\, \,\,\,\, q = \sqrt{(d - R)^2 + \zbh^2} \, ,
    \label{p_and_q}
    \end{split}
\end{equation}
respectively.

Following \cite{LassBlitzer:1983, Fukushima:2010} we can write the potential generated by the uniform ring as:

\beq
U_{\rm ring} =\frac{2 G M_{\rm ring}}{\pi p}  K(k)\,, \,\,\,\,\,\,\,\, k = \frac{4 R d}{p^2}\, ,
\eeq

and we can split the acceleration into one component perpendicular to the plane of the disk and aligned with $\zbh$ and another parallel to it in the direction of the cylindrical radius $d$: 
\begin{equation}
\mathbf{a}_{\rm ring} =a_d \,\mathbf{u}_d + a_z \, \mathbf{u}_z\, ,
    \label{eom_ring}
\end{equation}
where
$\mathbf{u}_d =\cos \phi_D \mathbf{u}_{x_{\rm BH}} + \sin \phi_D \mathbf{u}_{y_{\rm BH}}$ is the unit vector along $d$, $\mathbf{u}_z$ is the unit vector along $\zbh$ and 
\begin{align}
& a_d = - \frac{G M_{\rm ring} }{\pi p}\left[ \frac{K(k)}{d^2} + \left( 1 - \frac{z^2 + R^2}{d^2}\right) \frac{E(k)}{q^2}\right]  d,\label{Ad_disk}\\
& a_z = - \frac{2 G M_{\rm ring} }{\pi p q^2}  E(k) z\,, 
\label{Az_disk}
\end{align}
are the accelerations along $\mathbf{u}_d$ and $\mathbf{u}_z$, respectively.
$K(k)$ and $E(k)$ are the elliptic integrals of the first and second kind, respectively, and are defined in Appendix \ref{app:elliptic_int}.

The acceleration is singular when the point mass is on the equatorial plane ($z = 0$) and crosses the ring's radius ($d = R$), as a consequence of the argument of the elliptic integrals $k \rightarrow 1$ when approaching this singular point, leading to a divergence of $K(k)$ (while $E(k)$ remains always finite).

From Eq.~\eqref{Az_disk} it is clear that the acceleration along $z$ is always directed along $-\text{sgn}(z) \mathbf{u}_z$ and changes its sign accordingly, while the change of sign in $a_d$ is less trivial and it depends on how the values of $d$ and $z$ compare to $R$.

We can derive analytically the asymptotic limits of the acceleration defining $\epsilon = r/R$, when the point mass has a radius much smaller than the ring's size (or $\epsilon = R/r$ in the opposite regime) and expanding Eqs.\eqref{Ad_disk}-\eqref{Az_disk} at lowest order in $\epsilon$, finding  
\begin{align}
    & a_d^{\rm ring} \approx a_z^{\rm ring} \propto \frac{G M_{\rm ring}}{r^2}\, , \, & \text{if} \,\, R \ll r \, , \label{approx_uniform1} \\
    & a_d^{\rm ring} \approx a_z^{\rm ring} \propto \frac{G M_{\rm ring}}{2 R^3} r\, \, , \, & \text{if}  \,\, R \gg r \, .
    \label{approx_uniform2}
\end{align}

The limit in Eq.~\eqref{approx_uniform2} shows that rings located at $R> r$ also affect the motion of the point mass, generating an acceleration that goes as $r/R^3$, recovering the same scaling as that of the tidal acceleration due to a distant companion around a massive BH \citep{Cardoso:2021qqu}. This represents one important difference with a spherically symmetric distribution of matter, for which Gauss' theorem guaranties that any mass located at $R > r$ gives zero contribution to the total acceleration. 

\subsubsection{Power law disk}
\label{subsec:power_law}

Once that the single ring case is mastered, one should  consider that in a realistic scenario the matter density distribution over a disk is never perfectly uniform, but rather characterized by a dependency with the distance from the center. 

Specifically, we can consider that the surface density $\Sigma$ follows a power law
\begin{equation}
    \Sigma(r) =  \frac{B}{r^{\gamma}}\, ,
    \label{power_law_density}
\end{equation}
where $B$ is a constant.  

It has been shown that the power law in Eq.~\eqref{power_law_density} has exponent  $\gamma \sim 2$ when considering the CW stellar disk at the GC \citep{Paumard:2006, Bartko:2008ad}. However, this value of $\gamma$ is specific to this particular structure at the GC. More generally, we note the large-scale accretion flow at the GC is likely to follow a power law with index from $1$ to $3/2$~\citep{Gillessen:2018, Ressler:2018yhi, Cho:2023wqr}, the exact value depending on the magnetic field properties. To encompass these various cases, we consider values of $\gamma$ between $1$ and $2$. 

To integrate Eqs.~\eqref{Ad_disk}-\eqref{Az_disk} over a disk surface extending between $r'$ and $r' + dr'$, we consider an infinitesimal mass $\mathrm{d}M = 2 \pi \Sigma(r')r' \mathrm{d}r^{\prime}= 2 \pi B r'^{1-\gamma} \mathrm{d}r'$, such that the infinitesimal acceleration becomes:
\begin{align}
 \mathrm{d} a_d^{\rm disk} = - \frac{G \mathrm{d}M }{\pi p}\left[ \frac{K(k)}{d^2} + \left( 1 - \frac{z^2 + R^2}{d^2}\right) \frac{E(k)}{q^2}\right] d ,\label{Ap}
\end{align}
\begin{align}
 \mathrm{d} a_z^{\rm disk} = - \frac{2 G \mathrm{d} M }{\pi p q^2}  E(k) z\,.
\label{Aq}
\end{align}

The constant $B$ can be related with the total mass of the disk between $r_{\rm min}$ and $r_{\rm max}$ via
\begin{equation}
    M_{\rm tot}= 2 \pi\int_{r_{\rm min}}^{r_{\rm max}} B r^{\prime 1-\gamma}\mathrm{d}r^{\prime}  = 2\pi  
    \begin{cases}
         \frac{B(r_{\rm min}^{2 - \gamma} - r_{\rm max}^{2 - \gamma})}{\gamma -2} \,\,\text{for}\, \gamma \neq 2 ,  \\
        B \log \left(\frac{r_{\rm max}}{r_{\rm min}}\right) \,\, \text{for}\, \gamma = 2 .
     \end{cases}
    \label{Mtot}
\end{equation}

\noindent Finally, the acceleration given by the power-law density disk is
\begin{equation}
    \mathbf{a}_{\rm disk}= \int_{r_{\rm min}}^{r_{\rm max}} \mathrm{d} a_d^{\rm disk}\mathbf{u}_d + \mathrm{d} a_z^{\rm disk}\mathbf{u}_z\, .
\end{equation}

\begin{figure*}
   \centering
   \includegraphics[width=\textwidth]{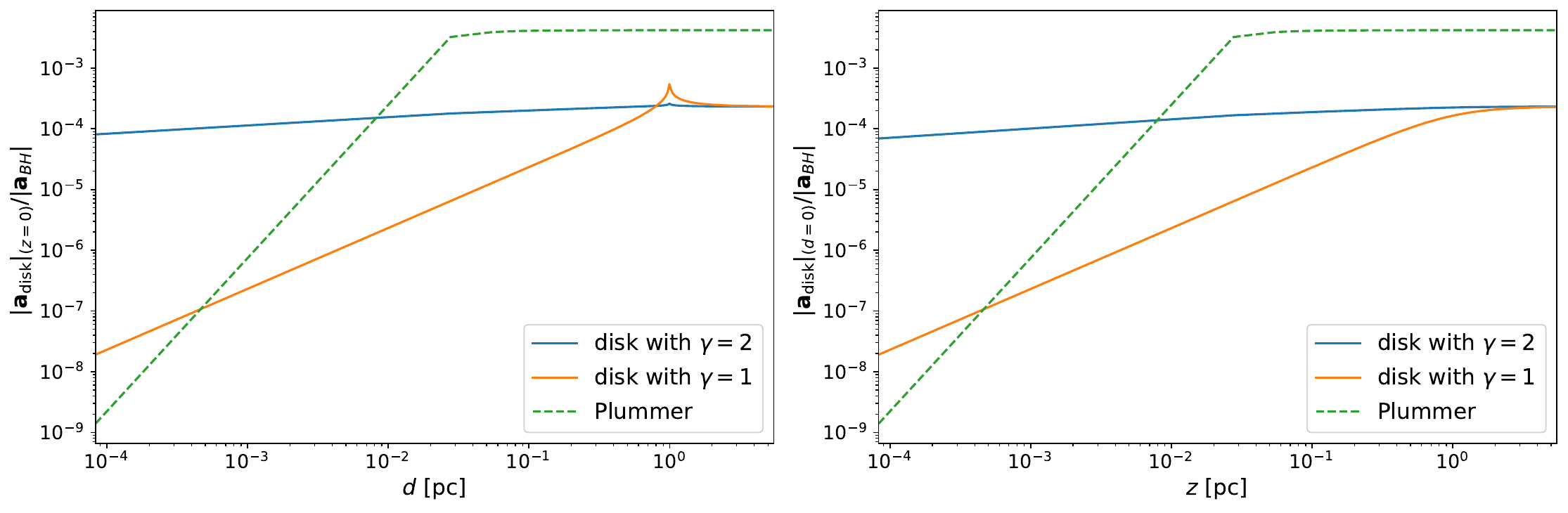}
      \caption{ Behaviour of the total acceleration $\mathbf{a}_{\rm disk}$ along the two directions of motion, for different values of $\gamma$, compared with the total (radial) acceleration due to Plummer's density profile, both normalized by the Newtonian acceleration induced by the SMBH. The total mass of the disk is $M_{\rm tot} = 10^3 M_{\odot}$, while $r_{\rm min} =10^{-6}\, \rm pc$ and $r_{\rm max} = 1 \, \rm pc$. For Plummer we use $r_0 = 0.012\, \rm pc$ and $\rho_0 = 1.69 \cdot 10^{-10}\, \rm kg/m^3$. In the left panel one can see the singularity at $d = r_{\rm max}$, while both components tend to the same limit as shown in Eqs.~\eqref{asymptotic_powerlaw1}-\eqref{asymptotic_powerlaw2}. The Plummer acceleration tends to a constant because of the normalization.}
    \label{fig:acc_tot}
\end{figure*}

The behaviour of $\mathbf{a}_{\rm disk}$ along the two directions, normalized by the Newtonian acceleration given by the SMBH, $|\mathbf{a}_{\rm BH}| = GM_{\bullet}/r^2$, is shown in Figure \ref{fig:acc_tot} for two values of $\gamma$ and a total mass of the disk $M_{\rm tot} = 10^3 \, M_{\odot}$. 

The disk acceleration is compared with the (radial) acceleration induced by the Plummer density profile, derived using $r_0 = 0.012 \, \rm pc$ and $\rho_0 = 1.69 \cdot 10^{-10} \, \rm kg/m^3$, i.e. the parameters used in e.g. \cite{Heissel:2021pcw}. Specifically, this value of $\rho_0$ is chosen such that the extended mass within S$2$'s apoastron corresponds to the current upper limit of the GRAVITY collaboration, $M(r_{\rm apo}^{S2}) \sim 10^{-3} \, M_{\bullet}$ \citep{GRAVITY:2021xju, GRAVITY:2024tth}. 

\begin{figure}
   \centering  
   \includegraphics[width=\hsize]{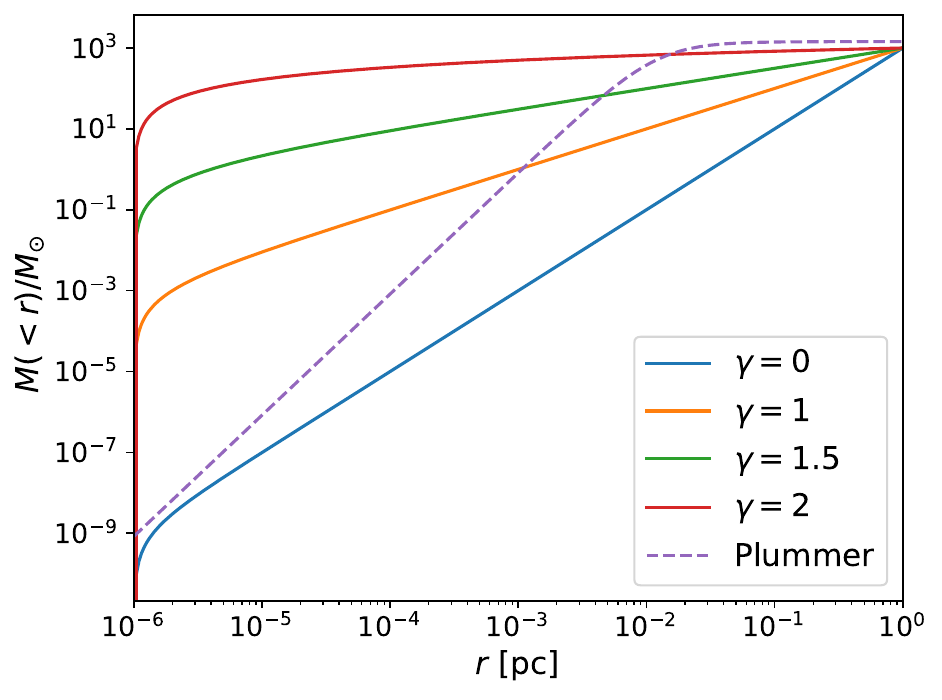}
      \caption{Behaviour of the mass distribution of the disk $M(<r)/M_{\odot}$ between $r_{\rm min } = 10^{-6}\, \rm pc$ and $r_{\rm max} = 1 \, \rm pc$, obtained using Eq.~\eqref{Mtot} and imposing a total disk's mass of $M_{\rm tot} = 10^3 \, M_{\odot}$, for different values of $\gamma$. The Plummer density is such that $M(<r_{\rm apo}^{S2}) \sim 10^{-3} \, M_{\bullet}$, which is slightly larger than the value chosen for the disk mass.}
         \label{fig:mass}
\end{figure}

The mass distribution $M(<r)$ for a total mass $M_{\rm tot} = 10^3 \, M_{\odot}$ is shown in Figure \ref{fig:mass} for different values of $\gamma$ and compared with the Plummer's mass described above.

The asymptotic behavior of $a_d$ and $a_z$ is given by
\begin{equation}
 a_d \approx a_z \propto \frac{G M_{\rm tot}}{r^2},  \,\,\,\,\,\,\,\,\,\,\, \text{if}\, r\gg  r_{\rm max} ,
    \label{asymptotic_powerlaw1}
\end{equation}
which means that we recover the Newtonian point-mass acceleration for any value of $\gamma$, while
\begin{equation}
       a_d \approx a_z \propto G M_{\rm tot} r \begin{cases}
      \frac{r_{\rm max}^3 - r_{\rm min}^3}{r_{\rm min}^3 r_{\rm max}^3 \log(r_{\rm max}/r_{\rm min})} \, \text{for} \, \gamma =2 ,\\
       \frac{r_{\rm max}^{1+\gamma} - r_{\rm min}^{1+\gamma}}{r_{\rm max}^{1 +\gamma}r_{\rm min}^3 - r_{\rm max}^3 r_{\rm min}^{1+\gamma}} \,\text{for} \, \gamma \neq 2,
      \end{cases} \text{if}\, r \ll r_{\rm min},
    \label{asymptotic_powerlaw2}
\end{equation}
where again we recover $a_d, a_z \propto r$. 

\subsection{Osculating formalism}

\subsubsection{Disk force in the Gaussian frame}
\label{subsec:gaussian_frame}
Once that the acceleration in the BH frame is known, one can compute the force in the comoving frame of the star, identified by the triad $\{\mathbf{n}, \boldsymbol{\lambda}, \mathbf{z}\}$, known as Gaussian frame, in order to determine how the osculating elements of the orbit are affected by the disk potential. 

To do so, we follow the standard perturbed $2$-body procedure, as explained in e.g. \cite{PoissonWill2012}, in which the force can be expressed as 
\beq
\mathbf{a_{\rm disk}} = \mathcal{R}_{\rm disk} \mathbf{n} + \mathcal{S}_{\rm disk}\boldsymbol{\lambda} + \mathcal{W}_{\rm disk}\mathbf{z}\, .
\label{perturbing_force}
\eeq
Since the three components of $\mathbf{a_{\rm disk}}$ must be evaluated in the Gaussian frame, one needs to apply a transformation matrix to the basis vectors $\{\mathbf{u}_{\xbh}, \mathbf{u}_{\ybh}, \mathbf{u}_{\zbh}\}$ to express them in terms of $\{\mathbf{n}, \boldsymbol{\lambda}, \mathbf{z}\}$. The procedure to obtain the acceleration in the Gaussian frame is reported in Appendix~\ref{app:transf_gauss}. 

In this frame, the three components of the force are given by
\begin{align}
 \mathcal{S}_{\rm disk} & = \mathbf{a}_{\rm disk} \cdot \boldsymbol{\lambda} \nonumber\\
 & =\frac{\sin \psi \cos \psi \sin^2 \ibh}{\left(\cos^2 \ibh \cos^2 \psi +\sin^2 \psi\right)^{1/2}} \int_{r_{\rm min}}^{r_{\rm max}} \mathrm{d}a_d \nonumber \\
 &- \sin \psi \sin \ibh\int_{r_{\rm min}}^{r_{\rm max}} \mathrm{d}a_z  ,
\label{S_disk}
\end{align}
\begin{align}
 \mathcal{W}_{\rm disk}& = \mathbf{a}_{\rm disk} \cdot \mathbf{z}  \nonumber \\
  &= - \frac{\cos \psi \sin \ibh \cos \ibh}{\left(\cos^2 \ibh \cos^2 \psi +\sin^2 \psi\right)^{1/2}} \int_{r_{\rm min}}^{r_{\rm max}}\mathrm{d}a_d \nonumber \\
  &+\cos \ibh  \int_{r_{\rm min}}^{r_{\rm max}} \mathrm{d}a_z   , 
\label{W_disk}
\end{align}
\begin{align}
    \mathcal{R}_{\rm disk} & =  \, \mathbf{a}_{\rm disk}\cdot \mathbf{n} \nonumber \\ 
&=\left(\cos^2 \ibh \cos^2 \psi +\sin^2 \psi\right)^{1/2}\int_{r_{\rm min}}^{r_{\rm max}} \mathrm{d} a_d \nonumber \\
&+ \cos \psi \sin \ibh \int_{r_{\rm min}}^{r_{\rm max}}\mathrm{d}a_z ,
\label{R_disk} 
\end{align}
where $\psi = \oorb + f - \beta$, $f$ being the true anomaly and $r =p_{\rm star}/(1 + e \cos f)$.


A clear difference between the disk's equations and those obtained from a spherically symmetric distribution is the presence of the two components \sdisk and \wdisk, which impact the evolution of the orbital elements.

The acceleration in Eqs.~\eqref{S_disk}-\eqref{W_disk}-\eqref{R_disk} depends on both the orbital parameters of the star and the disk's radial extension and orientation, and can consistently change in magnitude and direction according to the values chosen. 

Both components \sdisk and \wdisk show two points in which they change sign: these correspond to the intersection points between the plane of the orbit and the plane of the disk, i.e., the passage of the star through $\ell_{\rm BH}$, obtained solving $\zbh = 0$, which corresponds to $\psi = \pi/2$ and $f = \pi/2 + \beta - \oorb$. Due to the zero-thickness of the disk, this change of sign in $z$ produces a real discontinuity in \sdisk. In a realistic scenario, where the disk has some finite thickness, this change of sign would generally be smoother. 

In addition to that, \sdisk is zero in two other points, corresponding to $\psi = 0$, which give the passage thought $z_{\rm BH//orb}$. Here $\boldsymbol{\lambda}\perp\mathbf{\zbh}$ and since $\boldsymbol{\lambda} \perp \mathbf{n}$ by construction, this means that $\boldsymbol{\lambda} \perp \Pi_D$. In these points, \sdisk is continuous and the transition between positive and negative values is smooth.  

When comparing the disk effects to those induced by the $1$PN equations of motion we will consider 
\begin{align}
    & \mathcal{R} = \mathcal{R}_{\rm disk} + \mathcal{R}_{\rm 1PN} \, ,
    \label{Rtot}
    \\
   & \mathcal{S} = \mathcal{S}_{\rm disk} + \mathcal{S}_{\rm 1PN}\, ,
    \label{Stot}
    \\
    & \mathcal{W} = \mathcal{W}_{\rm disk}
\, ,\label{Wtot}
\end{align}
where $\mathcal{R}_{\rm 1PN}$ and $\mathcal{S}_{\rm 1PN}$ can be found in \cite{PoissonWill2012} and are reported in Appendix \ref{app:1PN} for completeness.

Instead, when considering higher order effects, i.e. the spin, we will compare \rdisk, \sdisk and \wdisk with the $1.5$PN effect only \citep{Dayem:2025cki}. 

We note that, contrary to $\mathcal{W}_{\rm 1.5 PN}$, \wdisk is zero both on the disk plane and orthogonally to it. This difference in the physical behavior between the disk and spin effects could have important observational consequences when disentangling the two effects is at stake. 

\subsubsection{Osculating equations for orbital elements}
\label{sec:orbital_elements}
The variation in time of the orbital elements of a point mass in a Keplerian orbit due to the presence of a perturbing acceleration is given by \citep{PoissonWill2012}
\begin{align}
 \frac{d p_{\rm star}}{dt}  = 2 \sqrt{\frac{p_{\rm star}^3}{G M_{\bullet}}}\frac{1}{(1-e \cos f)^3} \mathcal{S},
\label{dpdf} 
\end{align}

\begin{align}
\frac{de}{dt}  = \sqrt{\frac{p_{\rm star}}{G M_{\bullet}}} \left[ \sin f \, \mathcal{R} +  \frac{2 \cos f + e(1 + \cos^2 f)}{1 + e \cos f} \mathcal{S} \right] ,
\label{dedf}
\end{align}

\beq
\begin{split}
\frac{d \oorb}{dt}  = & \frac{1}{e} \sqrt{\frac{p_{\rm star}}{G M_{\bullet}}} \left[- \cos f\, \mathcal{R} + \frac{2 + e \cos f}{1 + e \cos f} \, \sin f \, \mathcal{S} \right. \\
& \left. - e \cot \iorb \frac{\sin(\oorb + f)}{1 + e \cos f} \mathcal{W} \right]\, ,
\label{domegadf}
\end{split}
\eeq
\begin{align}
\frac{d \iorb}{dt}  = \sqrt{\frac{p_{\rm star}}{G M_{\bullet}}} \frac{\cos(\oorb + f)}{1 + e \cos f} \mathcal{W} \, ,
\end{align}
\begin{align}
\frac{d \Oorb}{dt} = \sqrt{\frac{p_{\rm star}}{G M_{\bullet}}} \frac{\sin(\oorb + f)}{1 + e \cos f} \frac{\mathcal{W}}{\sin \iorb} \, .
\label{dOmegadf}
\end{align}

In deriving these expressions, no assumption has been made on the osculating force. This set of equations is thus valid for any disk's mass. To relate the above equations with the true anomaly $f$ one must consider the following equation
\begin{equation}
\begin{split}
    \frac{df}{dt} =& \sqrt{\frac{G M_{\bullet}}{p_{\rm star}^3}}(1 + e \cos f )^2 + \frac{1}{e}\sqrt{\frac{p_{\rm star}}{G M_{\bullet}}} \left[ \cos f \, \mathcal{R} \frac{}{} \right. \\
    &\left. - \frac{2 + e \cos f }{1+ e \cos f } \sin f \, \mathcal{S}\right]\, \\
    = & \left(\frac{df}{dt}\right)_{\rm Kepler} - \left(\frac{d\oorb}{dt} + \cos \iorb \frac{d \Oorb}{dt} \right)\, ,
    \end{split}
    \label{dfdt}
\end{equation}

such that the variation of the osculating element $\mu_i = \{p_{\rm star}, e, \oorb, \Oorb, \iorb\}$ can be computed as
\begin{equation}
    \mu_i (f) = \int_0^{T} \frac{d \mu_i}{dt} dt = \int_0^{f} \frac{d \mu_i}{dt} \frac{dt}{df} df \, .
    \label{average_change}
\end{equation}
In particular we can compute the secular variation over one orbit as $\Delta \mu_i = \mu_i(f) - \mu_{i}^0$, where $\mu_i^0$ are the initial parameters of the star.

\subsubsection{Observer-independent angles}
The orientation angles and in general the orbital parameters in Eqs.~\eqref{dpdf}-\eqref{dOmegadf} strongly depend on the choice of fundamental frame, as defined by \cite{PoissonWill2012}. 

Following \cite{Will:2023nlt, Dayem:2025cki}, we can define some observer-independent quantities that are invariant under rotations, namely

\begin{align}
&\frac{d \varpi}{dt} \equiv \frac{d \oorb}{dt} + \cos \iorb \frac{d \Oorb}{dt}\, , \label{dvarpidt}
\\
&\frac{d \Theta}{dt}  \equiv \sin \oorb \frac{d \iorb}{dt} - \sin \iorb \cos \oorb \frac{d \Oorb}{dt} \, , \label{dThetadt}
\\
& \frac{d \Xi}{dt} \equiv -\cos \oorb \frac{d \iorb}{dt} - \sin \iorb \sin \oorb \frac{d \Oorb}{dt} \,,\label{dxidt}
\end{align}
which solely depend on the orbital frame and can be used to compare different models regardless of the convention used. 

Specifically, $d \varpi/dt$ represents the frequency of the argument of periapsis shift, i.e., the in-plane precession, and it corresponds exactly to the non-Keplerian terms in Eq.~\eqref{dfdt}, while $ d \Theta/dt$ and $d \Xi/dt$ are the frequencies associated with the out-of-plane precession, i.e. the precession of the semi-major and semi-minor axis, respectively.  

For the LT precession, the variations over one orbit of $\varpi$ and $\Theta$ can be found analytically and they read \citep{Dayem:2025cki}
\begin{align}
    & \Delta \varpi_{\rm LT} = -8 \pi \left(\frac{GM_{\bullet}}{c^2 p_{\rm star}^2}\right)^{3/2} \chi \, \cos \theta \, ,\label{deltavarpi}\\
    & \Delta \Theta_{\rm LT} = -4 \pi \left(\frac{GM_{\bullet}}{c^2 p_{\rm star}^2}\right)^{3/2} \chi \sin \theta \sin \psi_{\rm BH} \,\, ,
    \label{deltaTheta}
\end{align}
where $\chi$ is the dimensionless spin of \sgra, $\theta$ plays the same role as $\ibh$ but with respect to the spin axis $\zbh$, which in general does not coincide with $z_{\rm disk}$, $\psi_{\rm BH} = \oorb + f -\beta_{\rm BH}$, where $\beta_{\rm BH}$ is defined in the same way as $\beta$ but with respect to the projection the BH spin axis over the orbital plane.

\section{Results}
\subsection{General power law disk}
In order to compute the variation of the elements in Eqs.~\eqref{dpdf}-\eqref{dOmegadf} and Eqs.~\eqref{dvarpidt}-\eqref{dThetadt}-\eqref{dxidt} we first consider star S$2$, whose motion is entirely described by the energy parameters
\begin{equation}
e = 0.8844, \,\,\,\,\,\,\,  p_{\rm star} = 0.0272 \, [\rm as]\, ,
\label{s2_energy_parameters}
\end{equation}
namely eccentricity and semi-latus rectum associated with the conservation of energy and angular momentum, the scale parameters, which are \sgra mass and GC distance $R_0$,   
\begin{equation}
     M_{\bullet}  = 4.29\,\, [10^6 \, M_{\odot}],\,\,\,\,\,\,\, R_0 =  8275.9\,  [\rm pc]\, ,
\label{GC_parameters}
\end{equation}
and the orientation angles with respect to the observer
\begin{equation}
\iorb = 134.67^\circ ,\,\,\,\,\,\,\, \oorb = 246.28^\circ,\,\,\,\,\,\,\, \Oorb =  48.21^\circ .
\label{s2_angles}
\end{equation}
An important point to note is that the orientation angles $\oorb$ and $\Oorb$ reported above do not coincide with $\omega_{\rm GRAV}$ and $\Omega_{\rm GRAV}$ in e.g., \cite{GRAVITY:2024tth} due to the different convention on the direction of $\zobs$. While $\iorb$ is unaffected by this axis change, $\oorb$ and $\Oorb$ must be changed according to
\begin{equation}
\oorb = \omega_{\rm GRAV} + \pi\,, \,\,\,\,\,\,\,\ \Oorb = \Omega_{\rm GRAV} + \pi\, .
    \label{angles_shifed}
\end{equation}
In this section, we focus on a hypothetical disk at the GC described by the power law density in  Eq.~\eqref{power_law_density} and total mass $M_{\rm tot} = 10^3 \, M_{\odot}$, a value chosen to reproduce the current spherically symmetric mass distribution upper limit, although this constraint cannot be directly applied to a disk-like structure.

Since the orientation angles of both systems are defined with respect to the same reference frame, the one of the observer, the relative inclination between the disk and the orbit can be found computing $\Delta \Omega = \Oorb - \Omega_{\rm disk}$ and replacing it in the expression 
 \begin{equation}
 \begin{split}
     \cos \ibh & = \hat{\mathbf{z}}_{\rm orb} \cdot \hat{\mathbf{z}}_{\rm disk} \\
     & = \sin i_{\rm disk} \sin \iorb \cos \Delta \Omega +\cos i_{\rm disk} \cos \iorb\,,
     \label{cosibh}
\end{split}
 \end{equation}
where $\hat{\mathbf{z}}_{\rm orb}$ and $\hat{\mathbf{z}}_{\rm disk}$ are the normal unit vectors to the orbital plane and the disk plane, respectively. 

Moreover, one can show (see Appendix~\ref{app:disk_vs_orbit}) that the angle $\beta$ can be related with the observer angles of both the star and the disk via
\begin{equation}
\begin{split}
    \cos \beta &=  - \sin i_{\rm disk} \sin \Delta \Omega\, . 
    \end{split}
\label{cosbeta}
\end{equation} 

Defining $r_{\rm peri} = p_{\rm star}/(1 + e)$ ($r_{\rm apo} = p_{\rm star}/(1 - e)$) as the point of closest (farthest) approach of the star to the SMBH, we observe that the maximum effect of the disk on $\Delta \mu_i$ is obtained for $r_{\rm min} = r_{\rm peri}$ and $r_{\rm max} = r_{\rm apo}$, as long as $r_{\rm min} \lesssim r_{\rm peri}$ and $r_{\rm max} \gtrsim r_{\rm apo}$. We thus focus on this configuration. 

\subsubsection{Secular trend of the semilatus rectum}
We start analysing the secular variation of $p_{\rm star}$. This only depends on $\mathcal{S}$, which includes both the disk and the $1$PN contribution. Contrary to the latter, which produces $\Delta p_{\rm star} = 0$, the disk induces a secular variation of the semi-latus rectum.

One can decompose the tangential acceleration \sdisk as 
\begin{equation}
\begin{split}
    \mathcal{S}_{\rm disk} &  \propto \mathcal{A}(f) \sin 2 \psi -\mathcal{B}(f) \sin \psi\\
     & \propto \mathcal{A}(f) (\sin 2 f \cos 2 \delta + \cos 2f \sin 2 \delta) \\
    &- \mathcal{B}(f) (\sin f \cos \delta + \cos f \sin \delta),
    \end{split}
 \end{equation}
where $\delta = \oorb - \beta$ and the terms $\mathcal{A}(f)$ and $\mathcal{B}(f)$ are defined as
\begin{align}
    & \mathcal{A}(f) =  \frac{ \sin^2 \ibh}{{\left(\cos^2 \ibh \cos^2 \psi +\sin^2 \psi\right)^{1/2}}} \int_{r_{\rm min}}^{r_{\rm max}}\mathrm{d}a_d (f), \\
    & \mathcal{B}(f) = \sin \ibh \int_{r_{\rm min}}^{r_{\rm max}} \mathrm{d} a_z(f) ,
\end{align}
and they depend on disk's parameters.

We note that spherically symmetric distributions induce zero secular variation of $p_{\rm star}$, and so do PN terms up to order $3$, which are well beyond the current observational reach. This means that the secular trend induced by the disk on $p_{\rm star}$ is a potentially important effect associated with a non-spherical extended mass. In the context of future studies where we will need to disentangle various effects, such specific features are important to stress.

Assuming that $p_{\rm star}(f)$ varies slowly and using only the Keplerian part of Eq.~\eqref{dfdt}, we get that $dp_{\rm star}/df \propto 1/(1 + e \cos f )^5$.
Since this term is even in $f$, all the terms multiplied by $\sin f $ or $\sin 2 f$ are null after integration over the orbit, and only those with $\cos f$ survive the average process.

Hence, the secular variation of $p_{\rm star}$ over one orbit is
\begin{equation}
    \Delta p_{\rm star} \propto  \sin 2 \delta \int \frac{\mathcal{A}(f) \cos 2f \mathrm{d}f}{(1 + e \cos f)^5}  - \sin \delta \int \frac{\mathcal{B}(f) \cos f \mathrm{d}f}{(1 + e \cos f)^5}.
    \label{deltapstar}
\end{equation}

When $\beta = \oorb$, i.e. when $z_{\rm disk//orb}$ coincides with the direction of periastron given by $\xorb$, $\delta = 0 $ and $\Delta p_{\rm star} = 0$, regardless of the other parameters of the disk.

The maximum and minimum of $\Delta p_{\rm star}$ in terms of $\beta$ can be obtained numerically deriving the expression above. We note that since \sdisk is invariant under the transformation $\beta \rightarrow \beta + \pi$, $\Delta p_{\rm star} (\beta) = \Delta p_{\rm star}(\beta + \pi)$.

\begin{figure}
   \centering
   \includegraphics[width=\hsize]{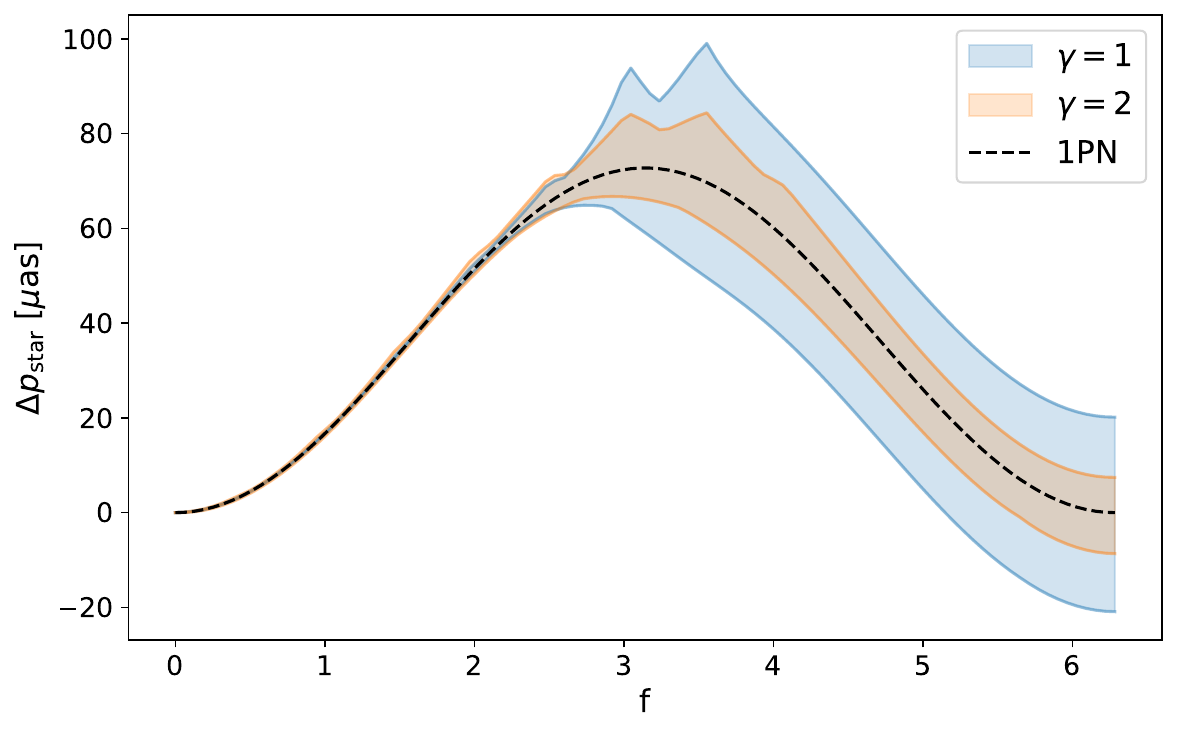}
      \caption{Variation of $p_{\rm star}$ over one orbit of S$2$ when $\ibh = \pi/2$, such that the disk effect is maximised in terms of inclination, and $M_{\rm tot} = 10^3 M_{\odot}$. The blue (orange) region represents the range of $\Delta p_{\rm star}$ with different $\beta \in [0, 2\pi)$ and different radial extent (as long as $r_{\rm min} \lesssim r_{\rm peri}$ and $r_{\rm max} \gtrsim r_{\rm apo}$) for $\gamma = 1$ ($\gamma =2$). Spikes represent the abrupt variation due to the passage through $z = 0$, which depends on the value of $\beta$. The black dashed line represents the variation in the $1$PN case that gives no secular variation. The fact that $p_{\rm star}$ varies secularly can be a distinctive feature of a disk-like distribution.}
         \label{fig:deltap}
\end{figure}

In Figure \ref{fig:deltap} we show the variation of $p_{\rm star}$ over one orbit with $\ibh = \pi/2$, because it maximises \sdisk. Compared to the value that minimizes it ($\ibh \sim 0$), there is roughly one order of magnitude difference in $\Delta p_{\rm star}$.
In this specific configuration, all the curves obtained for different values of $\beta \in [0, \pi]$, will fall in the blue (orange) region for $\gamma = 1$ ($\gamma = 2$).As already noted, provided we maintain $r_{\rm min} \lesssim r_{\rm peri}$ and $r_{\rm max} \gtrsim r_{\rm apo}$, varying the radial extent likewise causes the curves to fall within the coloured region. 

We note that the function $1/(1 + e \cos f)^5$ is maximal at apoapsis, and rapidly changes in this region. So, if the passage though $\zbh =0$ (that depends on $\beta$), which we have seen to lead to a discontinuity in the acceleration, happens in the vicinity of the apoastron passage, the contribution in $\Delta p_{\rm star}$ will also be maximum, leading to have a very sharp peak, while it is smoother if the passage occurs farther away from it . 

This effect is clearly enhanced for a thin disk like the one tested, as the passage though $z= 0$ represents a real singularity where the acceleration abruptly change sign. In a realistic scenario, where the disk has some finite thickness, the passage though $z=0$ will still be amplified if it happens near apoastron, but the peaks we observe in Figure~\ref{fig:deltap} will be smoother.

\subsubsection{In-plane precession and comparison with 1PN}
Regarding the in-plane precession encoded in $\Delta \oorb$, we note that, in general, it is maximised for co-planar orbits, since the star stays close to the dense region of matter for a longer time than highly inclined ones.

\begin{figure}
   \centering
   \includegraphics[width=\hsize]{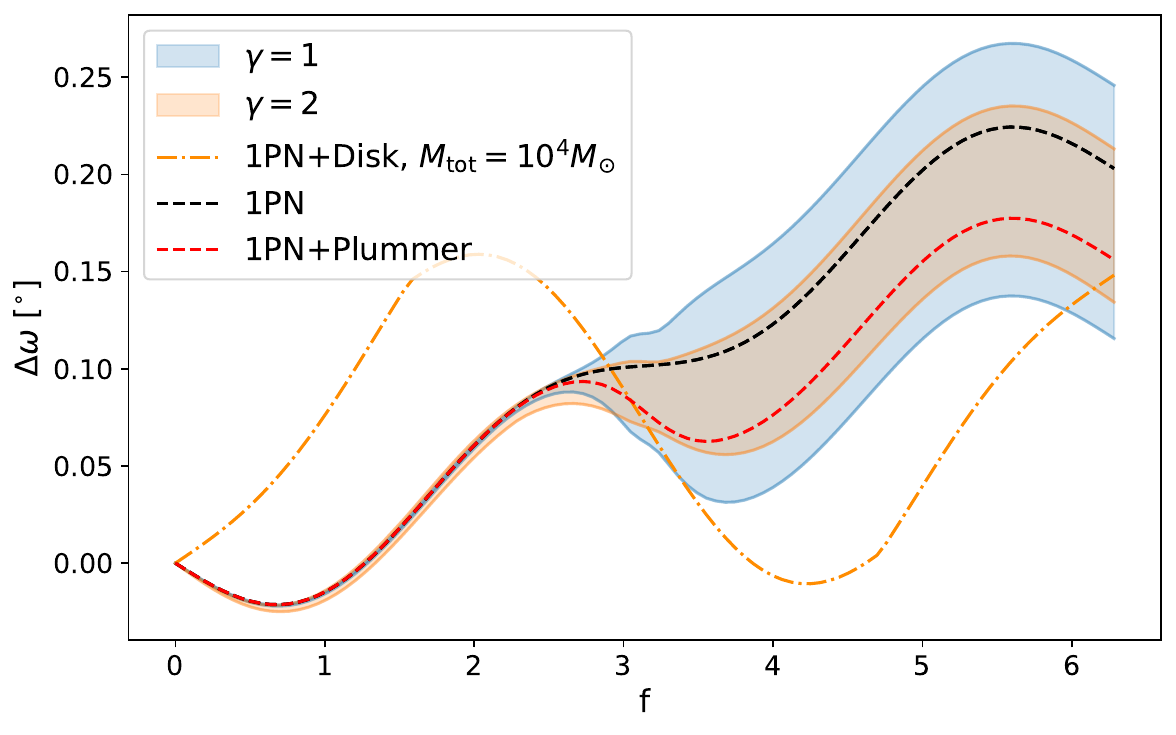}
      \caption{Variation of $\oorb$ over one orbit of S$2$ when $\ibh = 2^{\circ}$, such that the disk effect is is maximised in terms of inclination, and $M_{\rm tot} = 10^3 M_{\odot}$. The blue (orange) region represents the range of $\Delta \oorb$ with different $\beta \in [0, 2\pi)$ and different radial extent (as long as $r_{\rm min} \lesssim r_{\rm peri}$ and $r_{\rm max} \gtrsim r_{\rm apo}$) for $\gamma = 1$ ($\gamma =2)$. The black (red) dashed line represents the variation in the $1$PN case (1PN + Plummer). The dotted-dashed orange line represent the configuration with $\beta = \oorb$ and $M_{\rm tot} = 10^4 M_{\odot}$, to show that for specific orientations also larger masses are still compatible with current upper limits on the in-plane precession.}
         \label{fig:deltaomega}
\end{figure}

We thus focus on low inclination orbits, $\ibh = 2^{\circ}$, to maximise the disk's effect, since we cannot take the case with $\ibh = 0 $. The minimum shift in $\oorb$ is instead obtained when $\ibh = \pi/2$, and it is roughly one order of magnitude smaller. 
In Figure~\ref{fig:deltaomega} we show the variation $\Delta \oorb$ over one orbit, again taking the maximum and minimum shift, evaluated numerically, and showing in blue (orange) the region where all the curves would fall when changing $\beta \in [0, 2 \pi)$ and $\gamma =1$ ($\gamma = 2$).

For the chosen disk mass, the in-plane precession spanned by the disk models ($\gamma=1,2$) is comparable to that of a Plummer distribution (red dashed line, \citep{Heissel:2021pcw}), remaining close to the 1PN precession (black dashed line). However, we stress that this agreement is specific to this mass regime: for a disk mass of $M_{\rm tot} = 10^4 M_{\odot}$ and specific orientations, its precession can depart significantly from the Plummer case (cf. the dotted-dashed curve in Figure \ref{fig:deltaomega}), giving however the same total variation and being still compatible with the data. 

This means that flattened and spherical distributions can behave differently on the orbital motion, and it is thus necessary to constrain the disk model independently, rather than relying on the current limits obtained for spherically symmetric models. Future work will investigate this point in more depth.

\subsubsection{Out-of-plane precession and comparison with 1.5PN}

Although up to now we have assumed that \sgra is a Schwarzschild BH with the disk placed on its equatorial plane, the presence of an out-of-plane precession due to \wdisk, urges us to compare these effects with those induced by the LT precession.  

We start with the in-plane precession $\varpi$ in Eq.~\eqref{dvarpidt} and we compare the total shift due to the LT precession in Eq. \eqref{deltavarpi}, using $\chi = 1$ and $\theta = 0$ such that it is maximum, with the shift induced by the disk only.

It is important to underline that for nonzero spin, $\theta$ and $\ibh$ are different. Indeed, the latter encodes the inclination between the axis normal to the extended mass disk and the orbital angular momentum, while the former encodes the inclination between the spin axis and the orbital angular momentum. Only for zero spin we can safely assume that the two angles coincide, which boils down to assuming (without loss of generality) that the disk lies in the BH equatorial plane. 

Specifically, while it is possible to know $\ibh$ for some specific systems (e.g. the CW disk), $\theta$ is currently unconstrained.

As for $\Delta \oorb$, the effect of the disk is maximised for low inclination orbits, and thus we fix $\ibh = 2^{\circ}$, keeping in mind that our thin disk model is non defined exactly at zero inclination.

\begin{figure}
   \centering
   \includegraphics[width=\hsize]{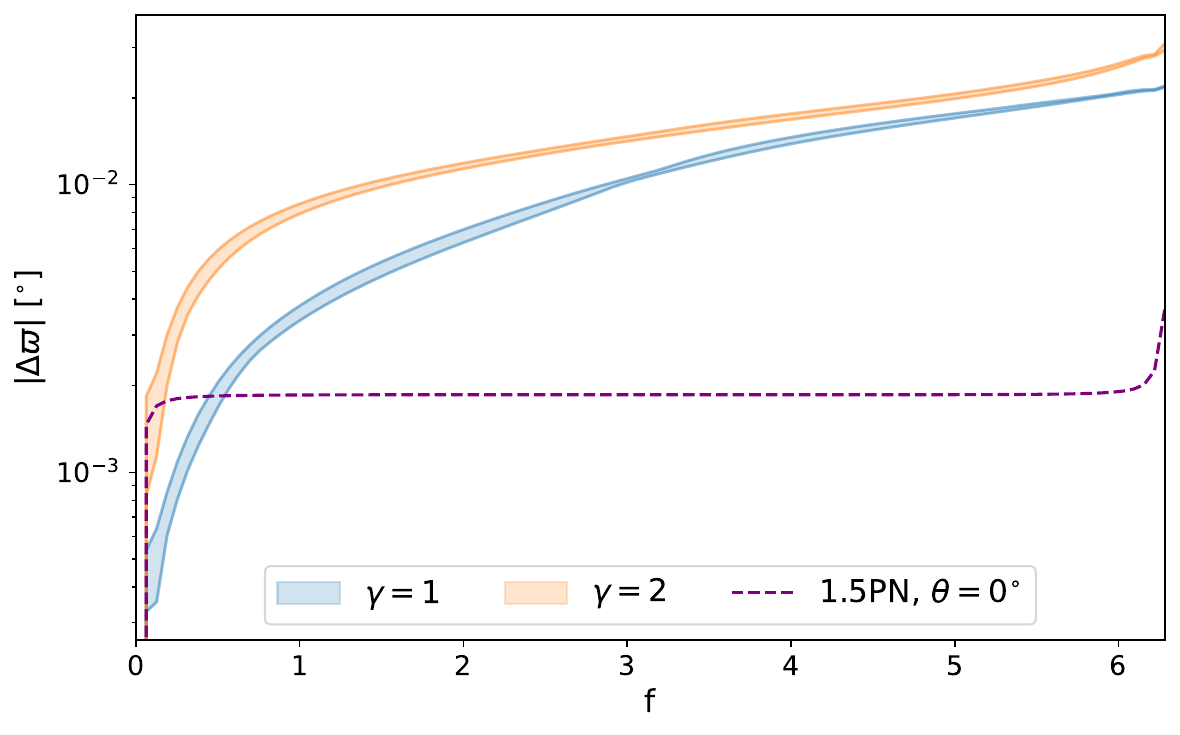}
      \caption{Absolute variation of $\varpi$ over one orbit of S$2$ when $\ibh = 2^{\circ}$, such that the disk effect is maximised in terms of inclination, and $M_{\rm tot} = 10^3 M_{\odot}$. The blue (orange) region represents the range of $\Delta \varpi$ with different $\beta \in [0, 2\pi)$ and different radial extent (as long as $r_{\rm min} \lesssim r_{\rm peri}$ and $r_{\rm max} \gtrsim r_{\rm apo}$) for $\gamma = 1$ ($\gamma =2$) considering only the disk. The purple dashed line represents the variation due to the spin obtained using the parameters that maximise $\Delta \varpi_{\rm LT}$ ($\chi =1, \theta = 0$).}
         \label{fig:deltavarpi_S2}
\end{figure}

From Figure~\ref{fig:deltavarpi_S2} one can see that the dependency on $\beta$ is basically washed out in $\Delta \varpi$. This follows directly from how $\varpi$ is defined in Eq.~\eqref{dvarpidt}. Indeed, whatever the inclination, the two terms depending on \wdisk in Eq.~\eqref{dvarpidt} cancel each other, and $\Delta \varpi$ depends on \rdisk and \sdisk only, which, in the specific case of almost equatorial orbits, weakly depend on $\beta$. 

We observe that the variation due to the $10^3 M_{\odot}$ disk is overall much larger than the one induced by the (maximum) LT in-plane precession, meaning that the variation of $\varpi$ can become an important feature to disentangle the spin and the disk effect in S$2$ motion. 

However, it is important to note that both effects are at least one order of magnitude smaller than the shift induced by the Schwarzschild precession, which generally dominates $\Delta \varpi$.

For the out-of-plane precession due to LT we take Eq.~\eqref{deltaTheta} with $\chi = 1$, $\theta = \pi/2$ and $\beta_{\rm BH} = \oorb - \pi/2$, such that $\Delta \Theta_{\rm LT}$ is maximised. 

\begin{figure}
   \centering
   \includegraphics[width=\hsize]{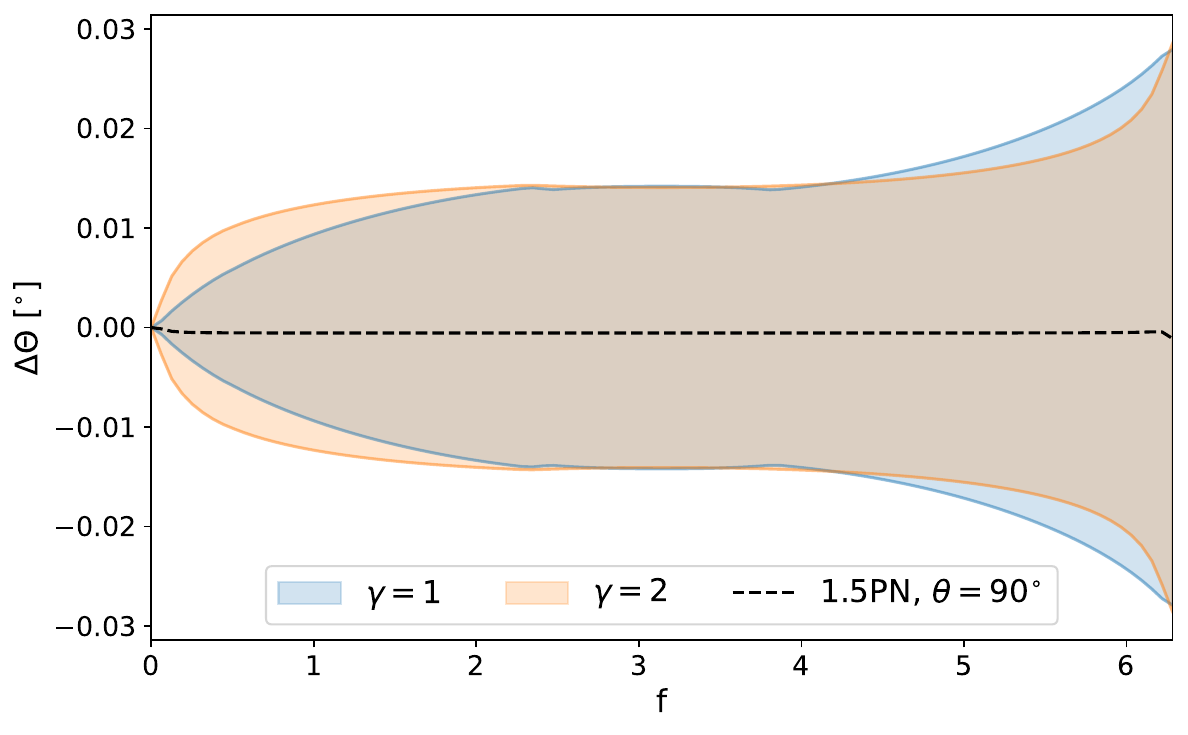}
      \caption{Variation of $\Theta$ over one orbit of S$2$ when $\ibh = 2^{\circ}$, such that the disk effect is maximised in terms of inclination, and $M_{\rm tot} = 10^3 M_{\odot}$. The blue (orange) region represents the range of $\Delta \Theta$ with different $\beta \in [0, 2\pi)$ and different radial extent (as long as $r_{\rm min} \lesssim r_{\rm peri}$ and $r_{\rm max} \gtrsim r_{\rm apo}$) for $\gamma = 1$ ($\gamma =2$) considering the disk only. The black dashed line represents the variation due to the spin obtained using the parameters that maximise $\Delta \Theta_{\rm LT}$ ($\chi = 1, \theta = 90^{\circ}$). For S$2$ the effect of the spin on the orbit is negligible, while the disk can produce significant out-of-plane precession, allowing us to constrain the disk mass.}
         \label{fig:deltatheta_S2}
\end{figure}

In the disk case, $\Delta \Theta$ is still maximised at low inclinations, so we fix again $\ibh = 2^{\circ}$. In Figure \ref{fig:deltatheta_S2} we show the maximum and minimum variation of $\Delta \Theta$. 

While the spin-induced shift is negligible in S$2$ motion, $\Delta \Theta_{\rm disk}$ is comparable to the current precision of our measurements of $\iorb$ and $\Oorb$, making it a real smoking gun for deriving upper limits on the disk mass, depending on its orientation and radial extent.  
\noindent We summarize the results obtained for S$2$ in Table~\ref{tab:shift_gen}.

The focus on the LT precession is motivated by the recent discovery of S$301$ by the GRAVITY+ collaboration \citep{GRAVITY:2026}. Unlike S$2$, this star is expected to be sensitive to the LT precession, opening the prospect of a measurement of the spin of \sgra within a decade. 

We therefore repeat our analysis using the current best-fit parameters of S$301$
 \citep{GRAVITY:2026}:
\begin{equation}
\begin{split}
& e_{\rm S301} = 0.98, \,\,\, p_{\rm S301} = 0.0033 \, \text{as}, \,\,\, i_{\rm S301} = 124^{\circ}\,(122^{\circ}), \\
& \Omega_{\rm S301} = 254^{\circ}\,   (77^{\circ}) ,\,\,\, \omega_{\rm S301} = 113^{\circ} \,(295^{\circ})\, ,
\end{split}
\end{equation}
where the angles are defined with respect to the observer (and again differ by $\pi$
from those in \cite{GRAVITY:2026}). The two possible values of $\omega_{\rm S301}$ and $\Omega_{\rm S301}$ (and consequently $i_{\rm S301}$), reflect the absence of radial velocity data, which prevents us from breaking the degeneracy between them. 

For the disk, we take $r_{\rm min}$ and $r_{\rm max}$ to be the periastron and apoastron of S$301$, while we keep the disk mass at $M_{\rm tot} = 10^3 \, M_{\odot}$.

\begin{figure}
   \centering
   \includegraphics[width=\hsize]{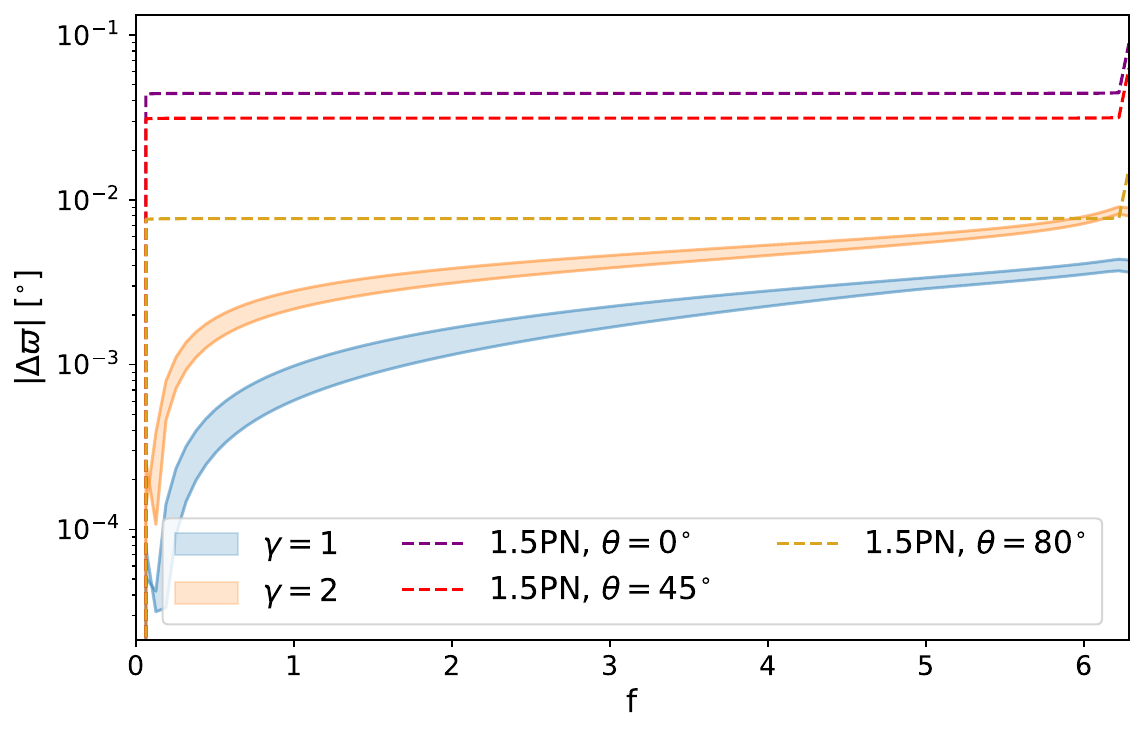}
      \caption{Absolute variation of $\varpi$ over one orbit of S$301$ when $\ibh = 2^{\circ}$, such that the disk effect is maximised in terms of inclination, and $M_{\rm tot} = 10^3 M_{\odot}$. The blue (orange) region represents the range of $\Delta \varpi$ with different $\beta \in [0, 2\pi)$ and different radial extent (as long as $r_{\rm min} \lesssim r_{\rm peri}$ and $r_{\rm max} \gtrsim r_{\rm apo}$) for $\gamma = 1$ ($\gamma =2$) considering only the disk. The dashed lines represent $\Delta \varpi_{\rm LT}$ for different values of the spin inclination $\theta$. }
         \label{fig:deltavarpi_S301}
\end{figure}

\begin{figure}
   \centering
   \includegraphics[width=\hsize]{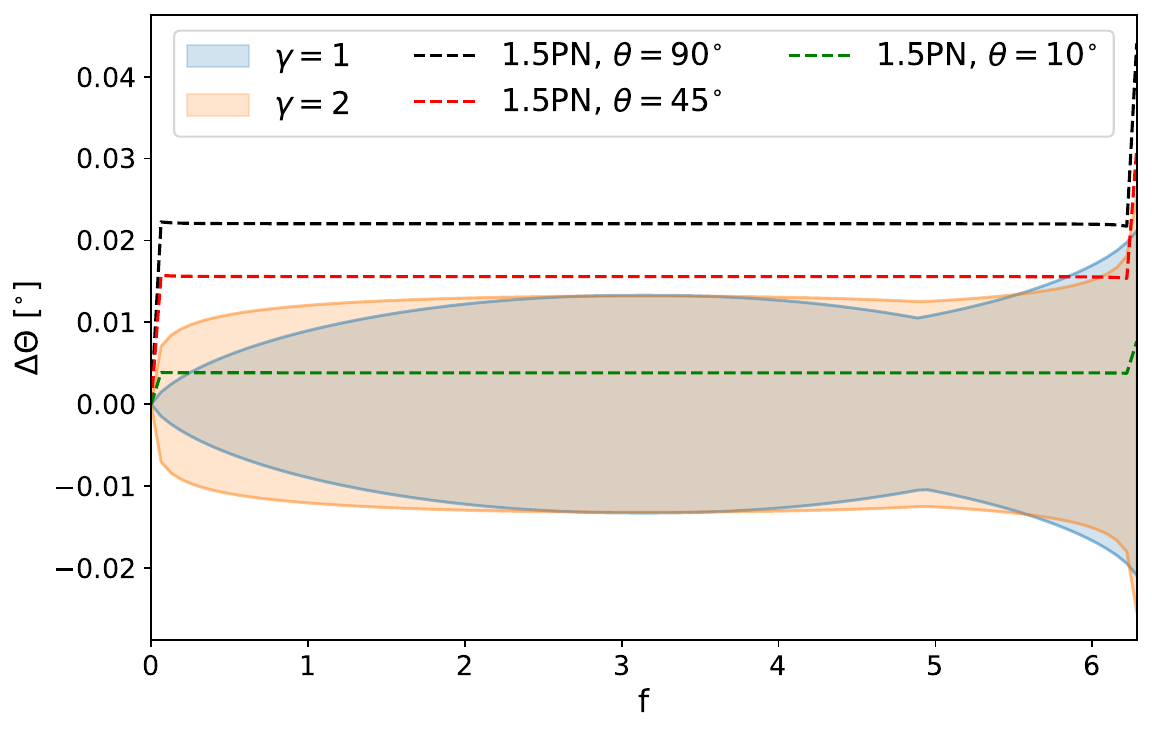}
      \caption{Variation of $\Theta$ over one orbit of S$301$ when $\ibh = 2^{\circ}$, such that the disk effect is maximised in terms of inclination, and $M_{\rm tot} = 10^3 M_{\odot}$. The blue (orange) region represents the range of $\Delta \varpi$ with different $\beta \in [0, 2\pi)$ and different radial extent (as long as $r_{\rm min} \lesssim r_{\rm peri}$ and $r_{\rm max} \gtrsim r_{\rm apo}$) for $\gamma = 1$ ($\gamma =2$) considering only the disk. The dashed lines represent $\Delta \Theta_{\rm LT}$ for different values of the spin inclination $\theta$. }
         \label{fig:deltatheta_S301}
\end{figure}

The resulting variations of $\varpi$ and $\Theta$ are shown in Figures \ref{fig:deltavarpi_S301} and \ref{fig:deltatheta_S301}. The disk contribution is comparable to that found for S$2$, but the spin effect varies by up to an order of magnitude as the spin-axis inclination $\theta$ is changed. Crucially, a value of $\theta$ that suppresses the effect on $\varpi$ enhances the effect on $\Theta$ and vice versa: this provides a means of breaking the degeneracy between the disk and spin contributions. Finally, the results scale linearly with $\chi$, which we set to its maximum value ($\chi = 1$).

While up to now our calculations are performed so that the effect of a hypothetical disk is maximised, in the next section we will focus on a more specific, realistic configuration. 
\begin{table}
\caption{Maxima shifts in S$2$ orbit for a disk with $M_{\rm tot} = 10^3 M_{\odot}$, and inclination such that its effect is maximised. Here we consider $r_{\rm min} = r_{\rm peri}$ and $r_{\rm max} = r_{\rm apo}$ (variations are smaller for $r_{\rm min}< r_{\rm peri}$ and $r_{\rm max}>r_{\rm apo}$) and the value of $\beta$ that also gives the maximum variation.}
\centering
\begin{tabular}{lccc}
\hline
X & $|\Delta X|$ & $\gamma$ \\ \hline
$ p_{\rm star, 1PN +disk}$ & $\sim 10 \, \mu \rm as$  & $1, 2$    \\
$ \omega_{\rm  1PN +disk}$ & $0.24^{\circ}$ & $1$    \\
$ \omega_{\rm  1PN +disk}$ & $0.21^{\circ}$ & $2$    \\
$ \omega_{\rm 1PN}$ & $0.20^{\circ}$ & -    \\ \hline
$\varpi_{\rm disk}$ & $\sim 0.02^{\circ}$ & $1, 2$  \\
$\varpi_{\rm LT}$ & $0.004^{\circ}$ &  -  \\
$ \Theta_{\rm disk}$ & $0.028^{\circ}$ & $1, 2$ \\
$ \Theta_{\rm LT}$ & $0.002^{\circ}$ & - \\
\hline
\end{tabular}
\label{tab:shift_gen}
\end{table}

\subsection{Stellar CW disk configuration}
\label{subsec:disk_config}
As already mentioned, the CW stellar disk at the GC is well described by the power law in Eq.~\eqref{power_law_density} with $\gamma =2$.

The orientation of the CW disk with respect to the observer is given by the angles:
\begin{equation}
    i_{\rm disk} = 127^{\circ} \pm 2^{\circ}\, , \,\,\,\,\,\, \Omega_{\rm disk} = 279^{\circ} \pm 2^{\circ}\, , 
\label{angles_CW_disk}
\end{equation}
where $\Omega_{\rm disk}$ differs by a factor $\pi$, as in \eqref{angles_shifed}, from the observed $\Omega$ reported in \cite{Paumard:2006}, while the radial extent of the CW disk is estimated to be between  
\begin{equation}
    r_{\rm min}^L \sim 0.04\, \text{pc} \, ,\,\,\,\,\,\,\,r_{\rm max}^L \sim 0.2-0.5\, \text{pc}\, \label{rmin_rmax}\, 
\end{equation}
\citep{Paumard:2006, Lu:2008iz, vonFellenberg:2022lyo, Goldwurm:2025}. 

These estimates for the radial extension of the CW disk are entirely dictated by luminous, observable matter. It therefore makes sense for the purpose of this work to consider also different inner and outer radii, as there are no constraints showing that dark remnants should follow the same extension as the luminous matter.

On the contrary, numerical simulations of the GC have shown that the radial mass segregation happening in spherically distributions is also found in axisymmetric structures, with heavier remnants concentrated towards the center \citep{Foote:2019icm}. Moreover, \cite{Panamarev:2022} showed that anisotropic vertical segregation drives heavier objects towards a thin, low inclination disk-like structure, reinforcing the flattening of the heavy components in a disk-dominated regime. 

We model the power law density of the CW disk as follows
\begin{equation}
    \Sigma = \frac{M_{\rm extra}}{2 \pi \log(r_{\rm min}^L/r_{\rm min}) r^2} + \frac{M_L}{2 \pi \log(r_{\rm max}^L/r_{\rm min}^L) r^2}\, ,
\end{equation}
i.e., we assume an extra mass $M_{\rm extra}$ distributed between $r_{\rm min}$ and the minimum luminous radius, and a luminous mass $M_L = 10^4 \, M_{\odot}$ distributed between $r_{\rm min}^L$ and $r_{\rm max}^L$ as derived by observations, with the same power law slope \citep{Paumard:2006}. 


\subsubsection{Application to S$2$}

Using Eqs.~\eqref{cosibh}-\eqref{cosbeta} we find that the relative orientation between S$2$ and the CW disk are
\begin{equation}
   \ibh^{\rm CW} \sim 86^\circ\, ,
   \,\,\,\,\,\,\,\,\,\beta^{\rm CW} \sim 232^{\circ}\, ,
\end{equation}

In Table~\ref{tab:shifts} we report the values of $\Delta \varpi_{\rm disk}$ and $\Delta \Theta_{\rm disk}$ for $M_{\rm extra} = 10^3 M_{\odot}$ and different values of the inner radius $r_{\rm min}$, where the case $r_{\rm min} = r_{\rm min}^L$ corresponds to $M_{\rm extra} = 0$. Since $r_{\rm max}^L > r_{\rm apo}$, varying the outer radius produces negligible variations and it is thus not considered. 


In the specific configuration of the CW disk we notice that the extra mass between $r_{\rm min}$ and $r_{\rm min}^L$ must be at least $M_{\rm extra} \gtrsim 5000 \, M_{\odot}$ in order to see a $\sim 20\%$ difference with the same quantities with $M_{\rm extra} = 0$.  

In addition to that, if we consider only the luminous radial extent of the CW disk, the mass compatible with S2 motion is up to $M_{\rm CW disk} \sim 10^6 \, M_{\odot} \sim 100 M_L$, meaning that there is space for a much larger mass than the luminous one, either if the disk coincides with the luminous radial extent, or it extends further inwards.

A proper fit of the non luminous mass contained in the CW disk, or other disk-like structures at the GC, is left for future works, since we have shown that S$2$ data can be used to constrain this parameter freely from any possible contamination from the spin. 

\subsubsection{Application to S$301$}

Applying the same calculation to S$301$, we find that the relative inclination between its orbital plane and the CW disk is given by
\begin{equation}
   \ibh^{\rm CW, S301} \sim 21^\circ\, (107^{\circ})\,,
   \,\,\,\,\,\,\,\,\,\beta^{\rm CW, S301} \sim 342^{\circ}\, (252^{\circ}) \, .
\end{equation}

Specifically, between the two possible orientations we take $\ibh = 21^{\circ}$ and $\beta=342^{\circ}$ because they maximise the disk's effect and we report the values of $\Delta \varpi_{\rm disk}$ and $\Delta \Theta_{\rm disk}$ in Table~\ref{tab:shifts}. 

In this configuration, $\Delta \Theta_{\rm disk}$ is roughly nine times smaller than the LT precession for S$301$.
However, we recall that we are considering the maximum spin configuration ($\chi = 1, \theta = \pi/2$), which means that it will likely be smaller, while the value of $M_{\rm extra}$ has not been determined yet, and could potentially be larger.

\begin{table}
\caption{Maxima shifts per orbit in the observables $\Delta \varpi$ and $\Delta \Theta$, using the CW disk configuration and different inner radii, while $r_{\rm max} = 0.5 \, \rm pc$ for both S$2$ and S$301$. The extra disk mass is $M_{\rm extra} =10^3 \, M_{\odot}$. When $r_{\rm min} = r_{\rm min}^L$, $M_{\rm extra} =0$, so the effect is due to the luminous stellar CW disk only. For completeness, we also report the in-plane Schwarzschild precession, the maximum LT precessions (from \cite{Dayem:2025cki}) and $\Delta \varpi = \Delta \oorb$ for a Plummer distribution (for S$2$ only).}\label{tab:shifts}
\centering
\begin{tabular}{lccc}
\hline
X & $|\Delta X|_{\rm S2}$ $[^{\circ}$] & $|\Delta X|_{\rm S301}$ $[^{\circ}$] & $r_{\rm min}$ \\ \hline
$ \varpi_{\rm disk}$ & $0.007$ & $0.004$ & $10^{-6}\, \rm pc$    \\
$\Theta_{\rm disk}$ & $5 \cdot 10^{-5}$ & $0.003$ & $10^{-6}\, \rm pc$ \\
\hline 
$ \varpi_{\rm disk}$ & $0.024$ & $0.012$ & $r_{\rm peri}^{\rm S2} (r_{\rm peri}^{\rm S301})$   \\[3pt]
$ \Theta_{\rm disk}$ & $0.0001$ & $0.005$ & $r_{\rm peri}^{\rm S2} (r_{\rm peri}^{\rm S301})$   \\[2pt]
\hline 
$ \varpi_{\rm disk}$ & $0.0002$ & $8 \cdot 10^{-6}$ & $r_{\rm min}^L$   \\[2pt]
$ \Theta_{\rm disk}$ & $1 \cdot 10^{-6}$ & $2 \cdot 10^{-6}$ & $r_{\rm min}^L$   \\[2pt]
\hline \hline
$ \varpi_{\rm LT}$ & $0.004$ & 0.089 &    \\
$ \Theta_{\rm LT}$ & $0.002$ & 0.044 & -  \\
\hline \hline
$\varpi_{\rm Sch}$ & $0.200$ & 1.684 &-   \\
$\varpi_{\rm Plum}$ & $0.044$ & - &-   \\
\hline
\end{tabular}
\label{tab:shifts}
\end{table}

\section{Conclusions}
In this work we analyze the effect that a thin disk-like structure at the GC would have on S2 and S301-like orbits. The setup is derived in a pure Newtonian framework, although the relativistic nature of \sgra is included via PN corrections.

We derive the components of the acceleration due to a power law density thin disk in the reference frame co-moving with the star, in order to determine the variation of the orbital elements due to the disk's perturbation.

In Section \ref{subsec:power_law} we consider the effect of a hypothetical disk on both S$2$ and S$301$ motion to show the variation induced on their orbital elements, assuming all possible orientations with respect to the observer and maximising its effect in terms of its inclination with respect to the orbital plane. 

For S$2$, we are interested in comparing $\Delta p_{\rm star}$ and $\Delta \oorb$ with the same quantities obtained in the $1$PN (and $1$PN + Plummer) framework, and we show that for specific orientation, the disk mass compatible with current observation of the in-plane precession can significantly differ from the same upper limits in the spherically symmetric case. 

The presence of \wdisk also forces us to compare the angles $\Delta \varpi$ and $\Delta \Theta$ with those obtained for the LT in-plane and out-of-plane precessions, showing that, while the spin effects on S$2$ are negligible, the disk induces variations that are compatible with the precision of the current measurements, becoming a smoking gun to disentangle the two effects and constrain the disk mass free from any degeneracy with the spin.

The same analysis is performed for S$301$, the newly discovered star which is sensitive to the LT precession. We found that while the disk contribution is comparable with that of S$2$, the spin effect can change significantly if different inclinations of the spin axis $\theta$ or different magnitudes $\chi$ are considered.

To apply the setup to a realistic scenario, we consider the stellar CW disk at the GC, such that its surface density, derived from the luminous matter, follows a power law with $\gamma =2$. 
In this way we are able to fix the orientation angles and study the evolution of the orbital elements of S$2$ and S$301$ varying the inner radius of the disk only, under the assumption that hypothetical dark mass aligned with the CW disk does not necessarily follow the same radial extend of the luminous mass, but actually segregate towards the center. 

For both stars, we compute the variations of the in-plane and out-of-plane precessions due to (maximum) LT and to the CW stellar disk, and we report them in  Table~\ref{tab:shifts}. 

With the configuration tested, i.e. assuming an extra mass of $M_{\rm extra} = 10^3 \, M_{\odot}$, the LT out-of-plane precession in S$301$ is roughly nine times larger than that induced by the CW disk. 

However, we note that the value of $\Delta \Theta_{\rm LT}$ is obtained for a face-on orbit ($\theta = \pi/2$) and a maximum spinning BH, while different limits have been obtained in the literature for \sgra spin, ranging from $0.1$ to $0.9$ \citep{Huang:2009,Moscibrodzka:2009, Broderick:2016,  Frangione:2020}, and the disk mass can potentially be larger than $10^3 \, M_{\odot}$. 
   
Hence, it is possible to derive upper limits on the disk mass using S$2$ motion, which is free from any spin bias, and derive them in terms of the possible orientations of the disk structure, starting from the CW disk.  

These upper limits can then be used to quantify the possible degeneracy, if any, between the LT and disk precessions in S$301$, and thus to obtain an unbiased measurement of the \sgra spin. The key point is that the spin-axis inclination $\theta$ affects $\Delta\varpi_{\rm LT}$ and $\Delta\Theta_{\rm LT}$ in opposite ways, whereas varying the disk inclination $\ibh$ shifts both in the same direction. This difference in behaviour opens up the possibility of cleanly disentangling the spin and disk effects.

Regarding the outer structures present at the GC we compute the acceleration due to the circumnuclear disk (CND) and the giant molecular cloud Sagittarius B$2$. 

The former has an inner radius $r_{\rm min} = 0.5 \, \rm pc$, an outer radius much less constrained at $r_{\rm max} = 3 - 7 \, \rm pc$ and a total mass of $M_{\rm tot} \sim 10^4 M_{\odot}$ \citep{Hsieh:2021, Solanki:2023mis}. To obtain a $0$-th order analysis we model the CND with a single ring at radius $R = r_{\rm min}$ with mass $M_{\rm tot}$.

The giant molecular cloud is instead modelled as a point mass with $M_{\rm SgrB2} \sim 10^7 \, M_{\odot}$ at a distance $R \sim 100\, \rm pc$ from \sgra \citep{2016A&A...588A.143S}. 

In this rough approximation that clearly overestimates the impact of both systems, the acceleration induced by both is roughly two orders of magnitude smaller than that induced by the luminous CW disk, and thus are both negligible. 

All the results published in this paper are reproducible using the Python package PERSEO (PERturbed Stellar Elements and Orbits), publicly available on GitHub \citep{foschi_perseo_2026}.

\begin{acknowledgements}
This work was supported by Paris Île-de-France Region; the French National Research Agency (ANR) under grant ANR-23-EDIR-0003 (GRAFITY), the "Action Thématique Gravitation Références Astronomie Métrologie" (ATGRAM), the "Action Thématique Phénomènes Extrêmes et Multimessagers" (ATPEM), and the "Action Thématique Cosmologie et Galaxies" (ATCG), of
CNRS/INSU, with co-funding by CNRS/IN2P3, CNRS/INP, CEA and CNES.
\end{acknowledgements}

\appendix

\section{Deriving the relation between $\oorb$ and $\obh$}
\label{app:omega}

\begin{figure}
   \centering
\includegraphics[width=\columnwidth]{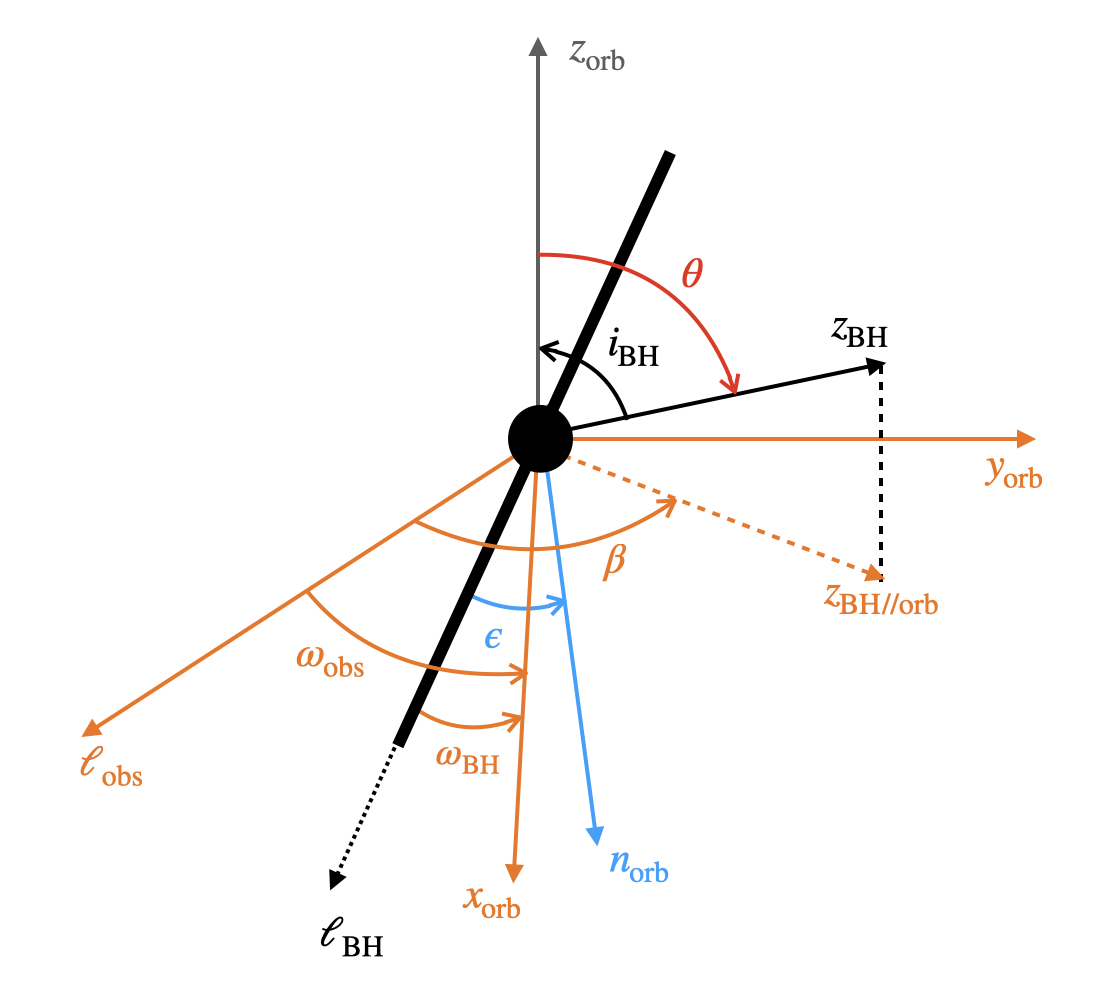}
      \caption{Schematic representation of the setup used to derive Eq.~\eqref{omega_bh}.}
         \label{fig:omega}
\end{figure}

Consider the schematic representation in Figure~\ref{fig:omega} where $\xorb$ is directed towards the pericenter passage of S$2$. As mentioned in the main text, $\ell_{\rm BH}$ represents the line of nodes between the BH frame and the orbit frame, while $\ell_{\rm obs}$ is the line of nodes between the orbit and the sky, i.e. the observer.

The projection of $\zbh$ over the orbital plane $(\xorb, \yorb)$ forms an angle $\beta$ with the line of nodes $\ell_{\rm obs}$, and we can express it in terms of the usual spherical angles $(\theta, \phi)$, with $\phi = \beta - \oorb$ and $\theta = \ibh$, the only difference being that they are defined in opposite directions,  
\begin{equation}
\begin{split}
    z_{\rm BH//orb} =& \sin \ibh \cos(\beta - \oorb) \xorb \\ &+ \sin \ibh \sin ( \beta - \oorb) \yorb  + \cos \ibh \zorb\, .
\end{split}
\end{equation}

With a similar reasoning we can write, 
\begin{equation}
    \ell_{\rm BH} = \cos \obh \xorb - \sin \obh \yorb\, .
\end{equation}

By definition, $\zbh \perp \ell_{\rm BH}$ and hence 
\begin{equation}
\begin{split}
    \zbh \cdot
    \ell_{\rm BH}& = \cos(\beta - \oorb) \cos \obh  -  \sin (\beta - \oorb) \sin \obh \\
    & = \cos(\beta - \oorb + \obh) = 0\, ,
\end{split}
\end{equation}
which is satisfied if $\beta - \oorb + \obh = \pm \pi/2$, corresponding to 
\begin{equation}
    \obh = \pm \frac{\pi}{2} - \beta + \oorb\, .
\end{equation}

To choose the sign of $\pm \pi/2$ one should consider that when the star passes the line of nodes $\ell_{\rm BH}$, $\zbh$ becomes positive.

One can write the normal unit vector as 
\begin{equation}
    \mathbf{n}_{\rm orb} = \cos( \epsilon-\obh) \xorb + \sin(\epsilon - \obh) \yorb\, ,
\end{equation}
where $\epsilon$ is the angle between the line of nodes $\ell_{\rm BH}$ and the position of the star identified by $\mathbf{n}_{\rm orb}$. 

If $\zbh$ must be positive for $\epsilon>0$, $\mathbf{\zbh} \cdot \mathbf{n}_{\rm orb}>0$, meaning  
\begin{equation}
\begin{split}
    \mathbf{\zbh} \cdot \mathbf{n}_{\rm orb} & = \cos(\epsilon - \obh) \cos (\beta - \oorb)  \\
    & +  \sin(\epsilon - \obh) \sin (\beta - \oorb) \\
    & = \cos(\epsilon - \obh - \beta + \oorb) \\
    &= \cos\left(\epsilon - s \frac{\pi}{2}\right) > 0\,.
\end{split}
\end{equation}
where $s = \pm 1$. The latter condition is satisfied for any $0 \leq \epsilon \leq \pi$ only if we pick $s = +1$. Hence, 
\begin{equation}
    \obh = \frac{\pi}{2} - \beta + \oorb\, .
\end{equation}

\section{Elliptic integrals}
\label{app:elliptic_int}

The elliptic integrals of first and second kind are defined, respectively, as
\begin{align}
    & K(k) \equiv \int_0^{\pi/2} \frac{d \theta}{\sqrt{1 - k \sin^2 \theta}}\\
    & E(k) \equiv \int_0^{\pi/2} \left(\sqrt{1 - k \sin^2 \theta}\right) d \theta \
\end{align}
where $k = 4 R d /p^2$.

\section{Transformation in the Gaussian frame}
\label{app:transf_gauss}
Using Eq.~$(3.40)$ of \cite{PoissonWill2012} we can express the fundamental frame $\{\mathbf{u}_{\xbh}, \mathbf{u}_{\ybh}, \mathbf{u}_{\zbh}\}$ in terms of the Gaussian frame $\{\mathbf{n}, \boldsymbol{\lambda}, \mathbf{z}\}$, getting 
\begin{align}
   \mathbf{u}_{\xbh} & = \cos \Obh \left( \mathbf{n} \cos \psibh - \boldsymbol{\lambda}\sin \psibh\right) + \sin \Obh \left(\mathbf{z} \sin \ibh \right. \nonumber \\
   & \left. - \cos \ibh(\boldsymbol{\lambda} \cos \psibh + \mathbf{n} \sin \psibh) \right)\, ,  \\
    \mathbf{u}_{\ybh} & = - \mathbf{z} \cos \Obh \sin \ibh + \sin \Obh \left( \mathbf{n} \cos \psibh - \boldsymbol{\lambda}\sin \psibh\right) \nonumber\\
   & + \cos \ibh \cos\Obh (\boldsymbol{\lambda} \cos \psibh + \mathbf{n} \sin \psibh) \, , \\
   \mathbf{u}_{\zbh} & = \mathbf{z} \cos \ibh + \sin \ibh \left( \boldsymbol{\lambda} \cos \psibh + \mathbf{n} \sin \psibh\right) \, ,
\end{align}
where $\psi_{\rm BH} = \obh + f$, $f$ being the true anomaly. In order to conclude the transformation, one needs to also express the coordinates $\{\xbh, \ybh, \zbh \}$ appearing in Eq.~\eqref{Ad_disk}-\eqref{Az_disk} as
\begin{align}
\xbh &= r (\cos \psi_{\rm BH} \cos \Obh - \cos \ibh \sin \Obh \sin \psi_{\rm BH})\, ,\label{xbh}\\ 
\ybh &= r (\sin \Obh \cos \psi_{\rm BH} + \cos \ibh \cos \Obh \sin \psi_{\rm BH})\, ,\label{ybh}\\ 
\zbh &= r \sin \ibh \sin \psi_{\rm BH}\, ,
\label{zbh}
\end{align}
with $r$ being now the radial distance of the star, that in the Keplerian two body problem can be written in terms of its eccentricity $e$ and semilatus rectum $p_{\rm star} = a(1 - e^2)$ (not to be confused with $p$ in Eq.\eqref{p_and_q}), that is $r = p_{\rm star}/(1 + e \cos f)$.

Replacing $\obh$ with Eq.~\eqref{omega_bh}, and defining $\psi = f - \beta + \oorb$ we get the acceleration in the comoving frame of the star:
\begin{align}
\begin{split}
    \mathbf{a}_{\rm disk}  = & \left[ \cos \psi  \left(A_z \sin \ibh + \cos \ibh (A_y \cos \Obh - A_x \sin \Obh) \right) \right.\\
    & -\left. \sin \psi (A_x \cos \Obh +  A_y \sin \Obh) \right] \mathbf{n} \\
    - & \left[ \sin \psi  \left(A_z \sin \ibh + \cos \ibh (A_y \cos \Obh - A_x \sin \Obh) \right) \right.\\
    &+\left. \cos \psi (A_x \cos \Obh +  A_y \sin \Obh) \right] \boldsymbol{\lambda}\\
    +& \left[ A_z \cos \ibh + \sin \ibh (A_x \sin \Obh - A_y \cos \Obh)\right] \mathbf{z}\, .
\end{split}
\end{align}
Finally, if we replace
\beq
A_x = a_d^{\rm disk} \xbh,\,\,\,\,\,\,\, A_y= a_d^{\rm disk} \ybh, \,\,\,\,\,\,\, A_z = a_z^{\rm disk} \zbh
\eeq
where $a_d$ and $a_z$ are the integrated accelerations in Eqs.\eqref{Ad_disk}-\eqref{Az_disk}, and using Eqs.~\eqref{xbh}-\eqref{zbh} we arrive to Eqs.~\eqref{S_disk}-\eqref{R_disk}.

\section{Osculating force due to the first Post Newtonian correction}
\label{app:1PN}
The three components of the osculating force when the first Post Newtonian correction is included in the equations of motion (cf. Eq.~\eqref{1pn}) are given by \citep{PoissonWill2012}:
\begin{align}
    & \mathcal{R}_{\rm 1PN} =  \frac{G^2 M_{\rm \bullet}^2}{c^2 p_{\rm star}^3} ( 1+e \cos f)^2 \left(3(e^2+1)+2 e \cos f  - 4 e^2 \cos^2 f
    \right),\\
     & \mathcal{S}_{\rm 1PN} = \frac{4 G^2 M_{\bullet}^2}{c^2 p_{\rm star}^3} \left(1 + e \cos f\right)^3 e \sin f,\\
      & \mathcal{W}_{\rm 1PN} = 0.
\end{align}

\section{Relative orientation between disk and orbit}
\label{app:disk_vs_orbit}
The normal unit vectors to the orbital plane and to the disk plane can be written in terms of the observed angles as
\begin{align}
    & \mathbf{\hat{z}}_{\rm orb} = \left(\sin \iorb \sin \Oorb, -\sin \iorb \cos \Oorb , \cos \iorb \right) \, ,\\
   &  \mathbf{\hat{z}}_{\rm disk} = \left( \sin i_{\rm disk} \sin \Omega_{\rm disk},  -\sin i_{\rm disk} \cos \Omega_{\rm disk} , \cos i_{\rm disk} \right)\, .
    \label{z_unit_vectors}
\end{align}

Since the two unit vectors are defined with respect to the same reference frame, the relative inclination between the two is simply given by 
\begin{align}
        \cos \ibh & = \mathbf{\hat{z}}_{\rm orb} \cdot \mathbf{\hat{z}}_{\rm disk} \nonumber \\
        & = \sin i_{\rm disk} \sin \iorb \sin \Omega_{\rm disk} \sin \Oorb \nonumber\\
       & +\sin i_{\rm disk} \sin \iorb \cos \Omega_{\rm disk} \cos \Oorb + \cos i_{\rm disk} \cos \iorb \nonumber\\
       & = \sin i_{\rm disk} \sin \iorb \cos \Delta \Omega +\cos i_{\rm disk} \cos \iorb\, , 
\end{align}
where $\Delta \Omega =\Oorb - \Omega_{\rm disk}$.

Following the same reasoning, the angle $\beta$ can be obtained from the scalar product
\begin{align}
    \cos \beta = & \boldsymbol{\ell}_{\rm obs} \cdot \mathbf{z}_{\rm disk//orb} \nonumber\\
     =& \boldsymbol{\ell}_{\rm obs} \cdot \mathbf{z_{\rm disk}} \nonumber\\
     =& \boldsymbol{\ell}_{\rm obs} \cdot \mathbf{z}_{\rm disk// obs}\, ,
\end{align}
where the first and second equalities come from the fact that $\boldsymbol{\ell}_{\rm obs} \perp \mathbf{\zorb}$ and $\boldsymbol{\ell}_{\rm obs} \perp \mathbf{\zobs}$, respectively.

The line of nodes in terms of the observer angles is  
\begin{equation}
    \boldsymbol{\ell}_{\rm obs } = \cos \Oorb \mathbf{\hat{x}}_{\rm obs} + \sin \Oorb \mathbf{\hat{y}}_{\rm obs} \, ,
\end{equation}
and using Eq.~\eqref{z_unit_vectors}, the scalar product reads
\begin{align}
    \cos \beta = & \boldsymbol{\ell}_{\rm obs} \cdot \mathbf{z}_{\rm disk// obs} \nonumber\\
     =& \sin i_{\rm disk} \left(\cos \Oorb \sin \Omega_{\rm disk} - \sin \Oorb \cos \Omega_{\rm disk}\right) \nonumber\\
     =& - \sin i_{\rm disk} \sin \Delta \Omega\, .
    \label{cos_beta}
\end{align}
Since $\beta$ is defined in $[0, 2\pi)$, Eq.~\eqref{cos_beta} is not sufficient to uniquely determine it. To do so we can compute 
\begin{align}
     \mathbf{z}_{\rm disk// orb} \cdot \mathbf{\xorb} &=\mathbf{z}_{\rm disk} \cdot \mathbf{\xorb}  \nonumber \\
     = &- \sin i_{\rm disk}(\cos \iorb \cos \Delta \Omega \sin \oorb + \cos \oorb \sin \Delta \Omega) \nonumber\\
     & + \cos i_{\rm disk} \sin \iorb \sin \oorb \nonumber\\
     & = \cos(\beta - \oorb)\, ,\label{cos_beta_omega}
\end{align}
and obtain $\sin \beta$ with a trigonometric expansion. In this way $\beta$ is defined unambiguously.

\bibliographystyle{aa} 
\bibliography{biblio}

\end{document}